\pdfoutput=1

\documentclass[twocolumn]{autart}    

\usepackage{graphicx}          
\usepackage[comma,numbers,square,sort&compress]{natbib}
\usepackage{epstopdf}
\usepackage{amsmath,amssymb,amsfonts}
\usepackage{bm}
\usepackage{textcomp}         
\usepackage{enumerate}   
\usepackage{color}      
\usepackage{subcaption}
\usepackage{algorithm}
\usepackage{algorithmicx}
\usepackage{makecell}
\usepackage{bibentry}
\nobibliography*

\usepackage{algpseudocode}  
\newcommand{\R}{{\mathbb R}}

\newcommand{\E}{{\mathbb E}}
\newcommand{\im}{{\mathrm i}}

\newtheorem{remark}{Remark} 
\DeclareMathOperator*{\trace}{trace}

\newtheorem{Assumption}{\bf Assumption}
\newtheorem{Proposition}{\bf Proposition}
\newtheorem{theorem}{\bf Theorem}
\newtheorem{lemma}{\bf Lemma}
\DeclareTextFontCommand{\texttt}{\ttfamily\upshape}

\def\pf{\textit{Proof. }}
\graphicspath{{./figures/}}
\begin{document}
	
	\begin{frontmatter}
		
		\title{An Efficient Implementation for Spatial-Temporal Gaussian Process Regression and Its Applications\thanksref{footnoteinfo}} 
		
		\thanks[footnoteinfo]{A preliminary version of this work \cite{ZKCLYZ20} was presented in the 39th Chinese Control Conference (CCC), 2020. Corresponding author Tianshi Chen. 	This work was supported by the Shenzhen Science and Technology Innovation Council under contract No. Ji-20170189 (JCYJ20170411102101881), the Robotic Discipline Development Fund (2016-1418) from Shenzhen Government, the general project funded by NSFC under contract No. 61773329, the Thousand Youth Talents Plan funded by the central government of China.} 
		
		\author[CUHKSZ]{Junpeng Zhang}\ead{junpengzhang@link.cuhk.edu.cn},    
		\author[CUHKSZ]{Yue Ju}\ead{yueju@link.cuhk.edu.cn}, 
		\author[AMSS]{Biqiang Mu}\ead{bqmu@amss.ac.cn},               
		\author[SYSU]{Renxin Zhong}\ead{zhrenxin@mail.sysu.edu.cn},
		\author[CUHKSZ]{Tianshi Chen}\ead{tschen@cuhk.edu.cn}               
		
		\address[CUHKSZ]{School of Data Science and Shenzhen Research Institute of Big Data, The Chinese University of Hong Kong, Shenzhen 518172, China} 
		\address[AMSS]{Key Laboratory of Systems and Control, Institute of Systems Science, Academy of Mathematics and Systems Science, Chinese Academy of Sciences, Beijing~100190, China}
		\address[SYSU]{School of Intelligent Systems Engineering, Sun Yat-sen University, Guangzhou, P.~R.~China}             

		
		\vspace*{-2mm}
		\begin{abstract}                          
			Spatial-temporal Gaussian process regression is a popular method for spatial-temporal data modeling. Its state-of-art implementation is based on the state-space model realization of the spatial-temporal Gaussian process and its corresponding Kalman filter and smoother, and has computational complexity $\mathcal{O}(NM^3)$, where $N$ and $M$ are the number of time instants and spatial input locations, respectively, and thus can only be applied to data with large $N$ but relatively small $M$. In this paper, our primary goal is to show that by exploring the Kronecker structure of the state-space model realization of the spatial-temporal Gaussian process, it is possible to further reduce the computational complexity to $\mathcal{O}(M^3+NM^2)$ and thus the proposed implementation can be applied to data with large $N$ and moderately large $M$. The proposed implementation is illustrated over applications in weather data prediction and spatially-distributed system identification. Our secondary goal is to design a kernel for both the Colorado precipitation data and the GHCN temperature data, such that while having more efficient implementation, better prediction performance can also be achieved than the state-of-art result. 
		\end{abstract}
		
	\end{frontmatter}
	
	\section{Introduction}\label{sec:introduction}
	
	Gaussian process regression is a popular method in statistical data modeling and analysis, closely related with the kernel method, e.g., \cite{HSS08}, and the kernel-based regularization method in system identification, e.g., \cite{PDCDL14}, and has wide applications in many fields such as machine learning, signal processing, and automatic control, e.g., \cite{WR06,PVMLS13,PDCDL14}. 
	In contrast with the common parametric modeling methods in system identification, e.g.,  the prediction error/maximum likelihood method \cite{Ljung:99}, its advantage lies in that, first, its model structure is determined by its covariance function (also called the kernel function), which incorporates the prior knowledge of the underlying function/system to be identified into the estimation procedure; second, its model complexity is governed by the parameter (called the hyper-parameter) used to parameterize the covariance function, and can be tuned in a continuous way. To apply Gaussian process regression methods, there are several issues that should be addressed, including the kernel design, e.g., \cite{Genton01,WA13,Chen18,CA21}, the hyper-parameter estimation, e.g., \cite{WR06,PDCDL14}, and the efficient implementation, e.g., \cite{QR05,BBBK11,CA21}. 
	Gaussian process regression has been used widely in dealing with the spatial-temporal data in many areas, such as climate science, social science, kriging, signal processing and physical inverse problems, e.g., \cite{PT04,GDGF10,AKK18,SSH13}. When dealing with the spatial-temporal data, the Gaussian process has two inputs: the locations and the time instants, and thus is often referred to as the spatial-temporal Gaussian process, e.g., \cite{SSH13,CTCSP16,TCCPS20}.

	For large scale spatial-temporal data, the aforementioned issues often become more involved. The current practice is to design a separable spatial-temporal kernel, which is a product of a spatial kernel and a temporal kernel, e.g., \cite{SSH13,CTCSP16,TCCPS20}, whose design should be based on the prior knowledge on the underlying function to be identified.  For a designed separable spatial-temporal kernel, many methods can be used for the hyper-parameter estimation, such as the empirical Bayes (EB) method (also called the marginal likelihood maximization (MLM) method), the Stein's unbiased risk estimate (SURE) minimization method, the generalized cross validation (GCV) method, e.g., \cite{WR06,PDCDL14}. The straightforward implementation of the hyper-parameter estimation and the following estimation and prediction step has computational complexity $\mathcal{O}(N^3M^3)$, where $N$ and $M$ are the numbers of temporal and spatial data, respectively, and thus is too expensive to be applied to large scale data. To reduce the computational complexity, it has been tried to first explore the structure of the temporal kernel, then derive a state-space model realization of the temporal Gaussian process in different ways, and finally convert the hyper-parameter estimation, function estimation and prediction to Kalman filtering, smoothing and prediction problems, e.g., \cite{SSH13,CTCSP16,KCYZ19,TCCPS20}. Such implementation has computational complexity $\mathcal{O}(NM^3)$ and thus is inefficient to be applied to spatial-temporal data with moderately large or large $M$. 
	
In this paper, we focus on the following two issues. Our primary focus is on the issue of how to further reduce the computational complexity such that the Gaussian process regression can be applied to spatial-temporal data with large $N$ and moderately large $M$. To tackle this problem, 
we first find that the state-space model realization of the spatial-temporal Gaussian process has a Kronecker structure and then 
by exploring this structure, we propose transformations for the original state-space model and then derive a new state-space model realization of the spatial-temporal Gaussian process. Finally, the Kalman filter, smoother and predictor are applied to handle the hyper-parameter estimation, function estimation and prediction, respectively. The proposed implementation is illustrated over applications in weather data prediction including the Colorado precipitation data considered in \cite{TCCPS20} and the Global Historical Climatology Network (GHCN) temperature data in \cite{GHCNdata,ZKCLYZ20},  and spatially-distributed system identification, e.g., \cite{QHJ18}.
Our secondary focus is to design a kernel for the two weather data sets, such that while having more efficient implementation, better prediction performance can also be achieved than the kernel proposed in \cite{TCCPS20}. To this purpose, the designed kernel should have state-space model realizations and also
incorporate the prior knowledge that both data sets are not strictly periodic but with slight temporal variation.

In contrast with the state-of-art result \cite{TCCPS20}, this paper has the following contributions: 
	\begin{itemize}
			\item[1)] a more efficient implementation algorithm with computational complexity $\mathcal{O}(M^3+NM^2)$ is proposed for the hyper-parameter estimation, the spatial-temporal Gaussian process regression and prediction, while the one in \cite{TCCPS20} has a computational complexity $\mathcal{O}(NM^3)$ and did not consider the efficient implementation of hyper-parameter estimation;

		\item[2)] a kernel is designed for the Colorado precipitation data and the GHCN temperature data and shown to give better prediction performance than the one in \cite{TCCPS20}.   

	\end{itemize}
Finally, in contrast with the preliminary version \cite{ZKCLYZ20} of this paper, we have included new theoretical results including Propositions \ref{prop:diagonal structure of Psi} to \ref{prop:computation of GCV and SURE cost function}, and Theorem \ref{thm:computational complexity}, designed a new kernel that gives better prediction performance for both the Colorado precipitation data and the GHCN temperature data, illustrated the implementation over a new application in spatially-distributed system identification, and included more implementation details, e.g., the derivation of the discrete-time state-space model realization of the temporal kernel, the treatment of the missing data and the selection of the starting points.

	The remaining parts of this paper are organized as follows. In Section \ref{sec:2GPR}, we first introduce some preliminary materials and then the problem statement. In Section \ref{sec:3NEW}, we propose an implementation with computational complexity $\mathcal{O}(M^3+NM^2)$. In Section \ref{sec:Case}, we test the proposed implementation over applications in weather data prediction and spatially-distributed system identification, where in Section \ref{sec:weather data}, we design a kernel and show its better prediction capability over the one in \cite{TCCPS20} for both the Colorado precipitation data and the GHCN temperature data. In Section \ref{sec:6Conclusion}, we give the conclusion of this paper. All proofs of theorems and propositions are included in Appendix A.

	\section{Preliminary and Problem Statement} \label{sec:2GPR}

	In this section, we first introduce some preliminary materials and then the problem statement of this paper. 
	

	\subsection{Spatial-temporal Function Estimation}
	
	In this paper, we consider the spatial-temporal function estimation problem described by
	\begin{align} \label{eq:NoisyObservation}
		y_{i,j} &= f(p_i,t_j) + v_{i,j} 
		,\\& \qquad i=1,\cdots,M, \ t_j=jT_s,j=1,\cdots,N+N_T,\nonumber
	\end{align}
	where $p_i\in\R^\nu$ with $\nu\in\mathbb{N}$ is the $i$th location, $t_j\in\R_+=\{x|x\geq0,x\in\R\}$ is the $j$th time instant, $f(p_i,t_j)$, $v_{i,j}\in\R$ and  $y_{i,j}\in\R$ are the unknown spatial-temporal function value, the measurement noise and the measurement output at the $i$th location and the $j$th time instant, respectively, $T_s>0$ the sampling interval, $M$ is the number of locations, and $N$ and $N_T$ are the numbers of time instants for the function estimation and validation, respectively.  The measurement noises $v_{i,j}$ with $i=1,\cdots,M$, $j=1,\cdots,N+N_T$ are assumed to be independently Gaussian distributed as follows
	\begin{align} \label{eq:noise}
		v_{i,j} \sim\mathcal{N}(0,\sigma^2).
	\end{align} 
	We aim to estimate the function $f:\R^m \times \R_+ \rightarrow \R$ based on the training data $\{p_i,t_j,y_{i,j}\}_{i=1,j=1}^{M,N}$
	such that it has as good prediction performance over the test data $\{p_i,t_j,y_{i,j}\}_{i=1,j=N+1}^{M,N+N_T}$ as possible.

	\subsection{Gaussian Process Regression}\label{subsec:GPR}
	
	Gaussian process regression  models the spatial-temporal function $f(p_i,t_j)$  as a spatial-temporal Gaussian process 
	\begin{align}
		\label{eq:stcovfunc}
		f(p_{i},t_{j})&\sim \mathcal{GP}(0,k(p_{i},t_{j},p_{i'},t_{j'};\alpha)),\\
		\label{eq:stcovfunc1}
		k(p_{i},t_{j},p_{i'},t_{j'};\alpha) 
		&=k_s(p_{i},p_{i'};\alpha_s)k_t(t_{j},t_{j'};\alpha_t),
	\end{align}
	where $i,i'=1,\cdots,M$, $j,j'=1,\cdots,N+N_T$, $\mathcal{GP}$ represents a Gaussian process, $k(p_{i},t_{j},p_{i'},t_{j'};\alpha)$ the covariance function (also called the kernel)  with a separable structure in space and time, e.g. \cite{SSH13,CTCSP16,TCCPS20}, $k_s(p_{i},p_{i'};\alpha_s):\R^{\nu} \times \R^{\nu} \rightarrow\R$ the spatial kernel, $k_t(t_{j},t_{j'};\alpha_t):\R_+ \times \R_+ \rightarrow\R$ the temporal kernel, $\alpha = [\alpha_t^T,\alpha_s^T ]^T \in \Omega \subset\R^d$ with $d\in\mathbb{N}$, $\alpha_s\in\R^{d_s}$ with $d_s\in\mathbb{N}$ and $\alpha_t\in\R^{d_t}$ with $d_t\in\mathbb{N}$ the hyper-parameters of $k$, $k_s$ and $k_t$, respectively, and $d =d_t+d_s$. It is assumed that 
	for any $i,i'=1,\cdots,M$, $j,j'=1,\cdots,N+N_T$, 	$ f(p_{i'},t_{j'})$ is independent of $v_{i,j}$.


The kernel $k(p_{i},t_{j},p_{i'},t_{j'};\alpha)$ determines the underlying model structure and its design for the two test data sets will be studied in Section \ref{sec:weather data}. The hyper-parameter $\alpha$ determines the model complexity and its estimation can be handled by many methods. Here,  
we consider the marginal likelihood  maximization (MLM) method, the generalized cross validation (GCV) method, and the Stein's unbiased risk estimation (SURE) method, e.g., e.g., \cite{PDCDL14}, which are listed below, respectively,
    \begin{align} \label{eq:loglikelihood}
    	\hat{\alpha}^{MLM} =& \mathop{\arg\min}_{\alpha\in\Omega}  \left\{ \frac{NM}{2}\log(2\pi) + \frac{1}{2}\log|\Sigma(\alpha)| \nonumber \right.\\  &\left. + \frac{1}{2}Y^T\Sigma^{-1}(\alpha)Y \right\},\\
    	\label{eq:GCV}
    	\hat{\alpha}^{GCV}=& \mathop{\arg\min}_{\alpha\in\Omega} \left\{ \frac{S}{NM(1-\delta/NM)^2}  \right\} ,\\
    	\label{eq:SURE}
    	\hat{\alpha}^{SURE} =& \mathop{\arg\min}_{\alpha\in\Omega} \{ S + 2\sigma^2\delta \},
    \end{align}
   where $\hat{\alpha}^{MLM}$, $\hat{\alpha}^{GCV}$ and $\hat{\alpha}^{SURE}$ denote the corresponding hyper-parameter estimate, 
    \begin{subequations}
    	\begin{align}
    		\label{eq:Sigma}
    		&\Sigma(\alpha) = K_t(\alpha_t)\otimes K_s(\alpha_s) + \sigma^2I_{NM}\in\R^{NM\times NM},   \\
    		&[K_t(\alpha_t)]_{jj'} = k_t(t_j,t_{j'};\alpha_t), \quad j,j' = 1,\cdots,N, \\
    		&[K_s(\alpha_s)]_{ii'} = k_s(p_i,p_{i'};\alpha_s), \quad i,i' = 1,\cdots,M, \label{eq:Ks} \\
    		&y_j = [y_{1,j},\cdots,y_{M,j}]^T\in\mathbb{R}^{M},  \label{eq:yj}\\
    		&Y =\left[ y_1^T, \cdots, y_{N}^T  \right]^T \in\mathbb{R}^{NM},\\
    		\label{eq:def of delta}
    		&\delta = \trace \left\{ \left[ K_t(\alpha_t)\otimes K_s(\alpha_s) \right] \Sigma(\alpha)^{-1} \right\}, \\
    		\label{eq:def of S}
    		&S = ||\hat{Y}-Y||^2_2,  \\
    		&\hat{Y} =\left[ K_t(\alpha_t)\otimes K_s(\alpha_s) \right]\Sigma(\alpha)^{-1}Y,
    	\end{align}
    \end{subequations}
$\otimes$ denotes the Kronecker product between two matrices, $I_{NM}\in\R^{NM \times NM}$ an $NM$-dimensional identity matrix, $[\ \cdot\ ]_{jj'}$  the $(j,j')$th entry of a matrix, $|\cdot|$ and $\trace(\cdot)$ the determinant and trace of a square matrix, respectively.

	\subsection{Problem Statement}
	
	
To state the problem, it is worth to note the following two observations.  Firstly, the state-of-art implementation in \cite{TCCPS20} designed a kernel such that	the spatial-temporal Gaussian process has a state-space model realization and then convert the function estimation problem to a Kalman filtering and smoothing problem, and the implementation has computational complexity $\mathcal{O}(NM^3)$, e.g., \cite{SSH13,CTCSP16,TCCPS20} and thus can be applied to data with large $N$ but relatively small $M$, e.g., the Colorado precipitation data with $N=1212$ and $M=367$ was studied in \cite{TCCPS20}. However, the implementation in \cite{TCCPS20} is still very expensive to apply for data with moderately large $M$, e.g., the GHCN temperature data with $N=6575$ and $M=3955$.  Secondly, the kernel designed in \cite{TCCPS20} does not give very good prediction performance for the Colorado precipitation data \cite{KCYZ19}, indicating there is a room to design better kernels.

The above observations motivate us to tackle the following two problems in this paper:
\begin{itemize}
	
			\item[1)] to develop implementation with lower computational complexity in terms of $M$ than the one in \cite{TCCPS20},  which can be applied to data with large $N$ and moderately large $M$, e.g., the GHCN temperature data;
		
		\item[2)] to design a kernel for both the Colorado precipitation data and the GHCN temperature data that gives better prediction performance than the one in \cite{TCCPS20}.

	\end{itemize}



	
	


	
	\section{An Efficient Implementation}\label{sec:3NEW}
	In this section, we propose a new implementation algorithm with computational complexity $\mathcal{O}(M^3+NM^2)$, which can thus be applied to data  with large $N$ and moderately large $M$. 
	
	\subsection{State-space Model Realization of Spatial-Temporal Gaussian Process}\label{subsec:state-space realization of spatial-temporal Gaussian process}
	
	
	For convenience, we assume in this section that the temporal kernel $k_t(t_{j},t_{j'};\alpha_t)$ is a stationary kernel  and then with a slight abuse of the notation, we can denote it by
	\begin{align} \label{eq:dtkernel}
		k_t(\tau;\alpha_t),\quad \tau=(j-j')T_s,\quad j,j'=1,\cdots,N+N_T.
	\end{align}
	Recall that the power spectral density (PSD) of a discrete-time kernel $k_t(\tau;\alpha_t)$, denoted as $\Phi(\omega)$, can be obtained by its discrete Fourier transform
	\begin{align}\label{eq:PSD}
		\Phi(\omega)=\sum_{\tau=-\infty}^{+\infty} k_t(\tau;\alpha_t)e^{-\mathrm{i}\omega\tau} \in\R_+, \mathrm{i} = \sqrt{-1}.
	\end{align}
	\begin{Assumption} \label{ass:timespacekernl}
		$\Phi(\omega)$ is a rational power spectral density with the order of $2r$ with $r\in\mathbb{N}$.
	\end{Assumption}
	Under Assumption \ref{ass:timespacekernl}, the spectral factorization technique, e.g., \cite{A12,GL00}, can be applied to \eqref{eq:PSD} and there exists a rational transfer function $W$ such that
	\begin{align}\label{eq:factorization}
		\Phi(\omega) = W(e^{\mathrm{i}\omega})W(e^{-\mathrm{i}\omega}).
	\end{align}
	From the realization theory of linear systems e.g., \cite{Chen99} and the transfer function $W(e^{\mathrm{i}\omega})$, for each location $p_i$, the corresponding discrete-time state-space model realization of a zero mean Gaussian process with the covariance function \eqref{eq:dtkernel} can be derived by
	\begin{align} \label{eq:StrictlyProperStateSpaceRepresentationd}
		\begin{aligned}
			{s}_{i,{j}} &= F_Ds_{i,{j-1}} + G_Dw_{i,{j-1}}, s_{i,0}\sim\mathcal{N}(0,\Sigma_{0}), \\
			z_{i,{j}} &= H_Ds_{i,{j}}, j= 1,\cdots,
		\end{aligned}
	\end{align}
	with $i,i'=1,\cdots,M$, $j,j'=1,\cdots,N+N_T$,
		\begin{align}
			\label{eq:Ezjj}
			\E[z_{i,{j'}}z_{i,j}] & =k_t(\tau;\alpha_t), 			\E[z_{i,j}z_{i',j}] =0, 
		\end{align}
	where $F_D\in\R^{r\times r}, G_D\in\R^{r}$ and $H_D\in\R^{1\times r}$ are the system matrix, the input matrix and the output matrix, respectively, $s_{i,{j}}\in\R^{r}$ is the state vector of the $i$th location at the $j$th time instant with $s_{i,0}$ and $s_{i',0}(i\neq i')$ being independent from each other, $w_{i,{j}}\in\R$ is white Gaussian noise with zero mean and unit variance, $\Sigma_{0}$ is the solution of the discrete-time Lyapunov equation $\Sigma_{0}=F_D\Sigma_{0}F_D^T+G_DG_D^T \in\R^{r\times r}$ and $\E(\cdot)$ is the mathematical expectation.
	
	Then we define that
	\begin{align}
		\label{eq:chi_j}
	\chi_j &=[f(p_1,t_j) ,\cdots,f(p_M,t_j)]^T\in\R^{M},
	\end{align}
and according to Assumption \ref{ass:timespacekernl} and \eqref{eq:stcovfunc1}, its covariance matrix is
	\begin{align}  \label{eq:Covfestimated}
		\begin{split}
			\E\left[ \chi_{j'}  \chi^T_{j}\right] =K_s(\alpha_s)k_t(\tau;\alpha_t)\in\R^{M\times M},
		\end{split}
	\end{align} where $K_s(\alpha_s)$ is defined in \eqref{eq:Ks}.	We let 
	\begin{align} \label{eq:zk}
		\begin{split}
			z_j = \left[z_{1,j}, \cdots, z_{M,j} \right]^T\in\R^M,
		\end{split}
	\end{align}
and then with \eqref{eq:Ezjj} and \eqref{eq:Covfestimated}, we obtain
	\begin{align} \label{eq:festimated}
		\begin{split}
			\chi_j = K_s(\alpha_s)^{1/2}z_j,
		\end{split} 
	\end{align}
	where $K_s(\alpha_s)^{1/2}$ is the ``square root'' of $K_s(\alpha_s)$ defined in \eqref{eq:SVDofKs}.
	With \eqref{eq:Covfestimated}-\eqref{eq:festimated}, we rewrite  \eqref{eq:StrictlyProperStateSpaceRepresentationd} as follows
	\begin{subequations} \label{eq:MatrixStateSpaceRepresentationd}
		\begin{align} \label{eq:MatrixStateSpaceStated}
			s_{j} &= F{s}_{j-1} + Gw_{j-1}, s_0\sim\mathcal{N}(0, I_M\otimes\Sigma_{0}),  \\
			\label{eq:MatrixStateSpaceOutputd}
			\chi_{j} &= H{s}_{j}, j = 1,2,\cdots,
		\end{align}
	\end{subequations}
	where $s_{j}=[s_{1,j}^T,\cdots,s_{M,j}^T]^T\in\R^{Mr}$, $F=I_M\otimes F_D\in\mathbb{R}^{Mr \times Mr}$, $G=I_M\otimes G_D\in\mathbb{R}^{Mr \times M}$, $H=K_s(\alpha_s)^{1/2}(I_M\otimes H_D)\in\mathbb{R}^{M \times Mr}$ and $w_{j}=[w_{1,{j}},\cdots,\\w_{M,{j}}]^T\in\R^{M}$.

	According to \eqref{eq:festimated} and \eqref{eq:Covfestimated}, the state-space model \eqref{eq:MatrixStateSpaceRepresentationd} is a realization of the Gaussian process \eqref{eq:stcovfunc}. Then the model \eqref{eq:NoisyObservation} can be accordingly rewritten as follows
	\begin{subequations} \label{eq:NewDTstate}
		\begin{align}
			\label{eq:ss1 for NewDTstate}
			s_{j } &= Fs_{j-1} + Gw_{j-1}, s_0\sim\mathcal{N}(0,I_M\otimes\Sigma_{0}), \\
			y_j &= Hs_j + v_j, j = 1,2,\cdots,
		\end{align}
	\end{subequations}
	where 
		$v_j=[v_{1,j},\cdots,v_{M,j}]^T\in\R^M$,
	with $v_{j}\sim\mathcal{N}(0,\sigma^2 I_M)$,  $y_j$ is defined in \eqref{eq:yj}, and $w_{j}$ and $v_{j}$ are independent for any $j=1,\cdots,N+N_T$.
	
	Then the spatial-temporal function estimation and prediction problem can be converted to a Kalman filtering, smoothing and prediction problem for \eqref{eq:NewDTstate} and the corresponding implementations has computational complexity $\mathcal{O}(NM^3)$, same as the ones in e.g., \cite{SSH13,CTCSP16,TCCPS20}.

	\begin{remark}\label{rmk:difference of {TCCPS20}}
		Note that Section \ref{subsec:state-space realization of spatial-temporal Gaussian process} and \cite[Proposition 2]{TCCPS20} use two different routes to derive the discrete-time state-space model realization  \eqref{eq:NewDTstate} of the spatial-temporal Gaussian process. It is not hard to show that they are equivalent in theory, but they are different in implementation. In particular, the discretization technique used in \cite{TCCPS20} includes solving an integral involving the matrix exponential, which needs to be handled carefully and if otherwise, numerical problem may occur, see e.g., \cite{WAG14} and the references therein. Therefore, the route in Section \ref{subsec:state-space realization of spatial-temporal Gaussian process} is preferable in practice, because no discretization of continuous-time state-space model is involved and thus possible numerical problems are avoided. Moreover, in practice one can design directly the discrete-time simulation-induced kernel \cite{Chen18} based on the prior knowledge, which is represented in a state-space model form.
		
\end{remark}

	\subsection{A Transformed State-space Model Realization}\label{subsec:transformed SSM}
	In order to further reduce the computational complexity in terms of $M$, it is useful to explore the Kronecker structure of the system, input and output matrices of the state-space model \eqref{eq:NewDTstate} and perform a coordinate and an output transformation to \eqref{eq:NewDTstate}.
	
	Firstly, we denote the singular value decomposition (SVD) of the spatial kernel matrix $K_s(\alpha_s)$ and its ``square root'' $K_s(\alpha_s)^{\frac{1}{2}}$ as follows
	\begin{align}
		\label{eq:SVDofKs}
		K_s(\alpha_s) &=\Lambda D \Lambda^T, K_s(\alpha_s)^{\frac{1}{2}} = \Lambda D^{\frac{1}{2}} \Lambda^T,
	\end{align}  
	where $D \in\mathbb{R}^{M \times M}$ is a diagonal matrix and its main diagonals are singular values of $K_s(\alpha_s)$, $D^{\frac{1}{2}}$ is a diagonal matrix with the square root of diagonals of $D$ and $\Lambda$ is an orthogonal matrix, i.e. $\Lambda\Lambda^T = \Lambda^T\Lambda = I_M$.
	
	Then for $j = 0,1,\cdots$, we introduce a state transform
	\begin{align}\nonumber
		x_{j} &= (\Lambda^T \otimes I_r) s_{j} \Longleftrightarrow s_{j}= (\Lambda \otimes I_r) x_{j} , \\
		\label{eq:state transformation}
		x_{j+1} 
		&=  (\Lambda^T \otimes I_r)(Fs_{j} +Gw_{j}) \nonumber\\
		& = (\Lambda^T \otimes I_r)F(\Lambda \otimes I_r)x_{j}+G\Lambda^Tw_{j} \nonumber\\
		& =  (I_M \otimes F_D) x_{j}+G\Lambda^Tw_{j},
	\end{align}
	where  $(\Lambda^T \otimes I_r)G =(\Lambda^T \otimes G_D)
	=(I_M \otimes G_D)(\Lambda^T \otimes 1)=G\Lambda^T $,
	and  an output transform for $j = 1,2,\cdots$,
	\begin{align} 
		\label{eq:lambdaT yj}
		l_j &= \Lambda^T y_j\\
		&=\Lambda^T (Hs_j+v_j) \nonumber\\
		&=\Lambda^TH(\Lambda \otimes I_r)x_{j} +\Lambda^T v_j \nonumber\\
		&=\Lambda^TK_s(\alpha_s)^{\frac{1}{2}}(I_M\otimes H_D) (\Lambda \otimes I_r)x_{j} +\Lambda^T v_j \nonumber\\
		&=D^{\frac{1}{2}}\Lambda^T(I_M\otimes H_D) (\Lambda \otimes I_r)x_{j} +\Lambda^T v_j \nonumber\\
		&=D^{\frac{1}{2}}(I_M\otimes H_D)x_{j} +\Lambda^T v_j,\nonumber
	\end{align}
    where the last equation is true because $\Lambda^{T}(I_{m}\otimes H_{D})(\Lambda\otimes I_{r})=(\Lambda^{T}\otimes 1)(\Lambda\otimes H_{D})=I_{M}\otimes H_{D}$.
	Then the state-space model \eqref{eq:NewDTstate} is transformed to
	\begin{subequations}
		\label{eq:NewSSM}
		\begin{align}
			\label{eq:stateupdate}
			x_{j } &= \bar{F} x_{j-1}+G\bar{w}_{j-1} , x_0 \sim\mathcal{N}(0,I_{M} \otimes \Sigma_0), \\ 
			\label{eq:l_j}
			l_j & = \bar{H}x_{j}+\bar{v}_j , j=1,2,\cdots,
			\end{align}
	\end{subequations}
	where $\bar{F}\in\mathbb{R}^{Mr \times Mr}$, $\bar{H} \in\mathbb{R}^{M \times Mr}$ and the covariance of $x_0$ are computed as
	\begin{subequations}
		\begin{align}
			\label{eq:F_bar H_bar}
			\bar{F} &=I_M \otimes F_D,\ \bar{H} = D^{\frac{1}{2}} (I_M\otimes H_D),\\
			I_{M} \otimes \Sigma_0&=(\Lambda \otimes I_r)(I_{M} \otimes \Sigma_0)(\Lambda^T \otimes I_r),
		\end{align}
	\end{subequations}
	$\bar{w}_j = \Lambda^T w_j  \sim\mathcal{N}(0,I_{M})$ and $\bar{v}_j = \Lambda^T v_j \sim\mathcal{N}(0,\sigma^2I_M)$. We denote the transformed output vector and its covariance matrix by $L$ and $\overline{\Sigma}(\alpha)$, respectively, which are described by
	\begin{subequations}\label{eq:relation2}
		\begin{align}
			\label{eq:Loutput}
			L =& [l_1^T, \cdots, l_N^T] =
			(I_N \otimes\Lambda^T) Y,\\
			\label{eq:def of Sigma_bar}
			\overline{\Sigma}(\alpha)
			=&\mathbb{COV}[L,L]
			=(I_N \otimes\Lambda^T)\Sigma(\alpha)(I_N \otimes\Lambda).   
		\end{align}
	\end{subequations}
	

	\subsection{Kalman Filter Based Estimation and Prediction}\label{subsec:KSKP}

	Firstly, we define the estimate $\hat{x}_{j|m}$ and its covariance matrix  $\overline{\Sigma}_{j|m}$ for $j=1,2,\cdots,m=0,1,\cdots,N$ as
	\begin{subequations}
		\label{eq:def of hat_xjm and overline_Sigmajm}
		\begin{align} \label{eq:hatx}
			\hat{x}_{j|m} &= \E[x_j|l_{0:m} ] ,\\
			\overline{\Sigma}_{j|m} &= \E[(x_j -\hat{x}_{j|m} )(x_j -\hat{x}_{j|m} )^T |l_{0:m} ],
		\end{align}
	\end{subequations}
	where $l_{0}$ is a null vector and
	\begin{align}\label{eq:output transform sequence}
		l_{0:m} = \left\{l_{0},\cdots,l_{m}  \right\}.
	\end{align}
	Then the Kalman filter for  \eqref{eq:NewSSM} can be expressed as 
	\begin{subequations}\label{eq:NewKalmanFilter}
		\begin{align}
			\label{eq:Nek}
			\bar{e}_j  &= l_j -\bar{H}\hat{x}_{j|j-1} , \\
			\label{eq:def of E_bar_j}
			\bar{E}_j & =\mathbb{COV}[\bar{e}_j,\bar{e}_j] \\
			\label{eq:NEk}
			&= \bar{H} \overline{\Sigma}_{j|j-1}\bar{H}^T + \sigma^2I_M , \\
			\label{eq:Nxkk}
			\hat{x}_{j|j} &= \hat{x}_{j|j-1} +\overline{\Sigma}_{j|j-1}\bar{H}^T\bar{E}_j^{-1}\bar{e}_j,\\
			\label{eq:NSigmakk}
			\overline{\Sigma}_{j|j} &= \overline{\Sigma}_{j|j-1} - \overline{\Sigma}_{j|j-1} \bar{H}^T \bar{E}_j^{-1}\bar{H}\overline{\Sigma}_{j|j-1},\\
			\label{eq:Nxk}
			\hat{x}_{j+1|j} &= \bar{F} \hat{x}_{j|j},\\
			\label{eq:NSigma}
			\overline{\Sigma}_{j+1|j} &= \bar{F} \overline{\Sigma}_{j|j}  \bar{F}^T + Q,
		\end{align}
	\end{subequations} 
	where $\bar{e}_j$ is known as the innovation, 
	\begin{align}\label{eq:def of Q}
		Q=I_M\otimes(G_DG_D^T)
	\end{align}
    and the iterative algorithm starts from $j=1$.

	For the purpose of function estimation, we apply Kalman smoother as follows
	\begin{subequations}\label{eq:NewKalmanSmoother}
		\begin{align}
			\label{eq:NSigmajj}
			\overline{\Sigma}_{j|j} &= \overline{\Sigma}_{j|j-1}- \overline{\Sigma}_{j|j-1} \bar{H}^T \bar{E}_j^{-1}\bar{H}\overline{\Sigma}_{j|j-1}, \\
			\label{eq:barJ1}
			\bar{J}_j &= \overline{\Sigma}_{j|j}\bar{F}^T\overline{\Sigma}_{j+1|j}^{-1}, \\
			\label{eq:hatx1}
			\hat{x}_{j|N} &= \hat{x}_{j|j} + \bar{J}_j(\hat{x}_{j+1|N}-\bar{F}\hat{x}_{j|j}), \\
			\label{eq:NSigmakN1}
			\overline{\Sigma}_{j|N} & =\overline{\Sigma}_{j|j} 
			+ \bar{J}_j(\overline{\Sigma}_{j+1|N}
			-\overline{\Sigma}_{j+1|j})\bar{J}_j^T, \\
			\label{eq:NfkN1}
			\hat{f}_{j|N} &= \Lambda  \bar{H}\hat{x}_{j|N} 
			,\ j=N-1,\cdots,1,
		\end{align}
	\end{subequations}
	where $\hat{f}_{j|N} = \E \left[ \chi_j | l_{0:N}  \right]$ with $ \chi_j $ defined in \eqref{eq:chi_j}.

	For the purpose of function prediction, we apply the Kalman predictor as follows 
	\begin{subequations} \label{eq:NewKalmanPredictor}
		\begin{align}
			\label{eq:hatx2}
			\hat{x}_{j|N} &= \bar{F}\hat{x}_{j-1|N}, \\
			\label{eq:NSigmakN3}
			\overline{\Sigma}_{j|N} &= \bar{F}\overline{\Sigma}_{j-1|N}\bar{F}^T + Q, \\
			\label{eq:hatf2}
			\hat{f}_{j|N} &= \Lambda \bar{H}\hat{x}_{j|N}  
			,\ j=N+1,\cdots,N+N_T,
		\end{align}
	\end{subequations}
	where $\hat{f}_{j|N} = \E \left[ \chi_j | l_{0:N}  \right]$ is the prediction of $ \chi_j $ at $j$th time instant. 
	
	
	
	\subsection{Hyper-parameter Estimation} \label{subsec:HypEst}
	Based on the Kalman filter \eqref{eq:NewKalmanFilter}, it is possible to propose efficient implementation algorithms for the MLM, GCV and SURE methods.

	\begin{lemma}\cite[p.~302, Properties of the Innovation Sequence]{Candy05}\label{lemma:preliminary results of the innovation sequence}
		 For $j=1,\cdots,N$, the innovation $\bar{e}_{j}$ in \eqref{eq:Nek} can be represented as a linear function of $l_{0:j}$ in \eqref{eq:output transform sequence}, i.e.
		 \begin{subequations}
		 	\begin{align}
		 		\label{eq:linear representation form of innovation for j=1}
		 			&\bar{e}_{1}=l_{1},\\
		 			\label{eq:linear representation form of innovation for j>1}
		 			&\bar{e}_{j}=l_{j}-\sum_{i = 1}^{j-1} b_{j,i} l_{i},\ \text{for}\ j=2,\cdots,N,\\
		 			 \label{eq:uncorrelatedness of innovation}
		 			 	&\mathbb{COV}[\bar{e}_{j},\bar{e}_{j'}]=0,\ \text{for}\ j'=1,\cdots,N\ \text{and}\ j\neq j',
	 	    \end{align}
		 \end{subequations}
 	      where $b_{j,i}\in\R$ is the corresponding coefficient for $i = 1,\cdots,j-1$.
	\end{lemma}

	\begin{Proposition}\label{prop:diagonal structure of Psi}
		Let
		\begin{subequations}\label{eq:Theta and Psi}
			\begin{align}\label{eq:def of Theta}
				\Theta =& [\bar{e}_1^T,\cdots,\bar{e}^T_N]^T,\\
				\label{eq:def of Psi}
				\Psi = &\mathbb{COV}[\Theta,\Theta],
			\end{align} 
		\end{subequations}
	where $\bar{e}_j$, $j=1,\cdots,N$, are defined in \eqref{eq:Nek}. Then following Lemma \ref{lemma:preliminary results of the innovation sequence},
		$\Theta$ and $\Psi$ can be rewritten as
		\begin{subequations}	\label{eq:proposition1}	
			\begin{align}
				\label{eq:equality of Theta and Gamma}
				\Theta &= \Gamma L,\\
				\label{eq:Psi using Gamma}
				\Psi &= \Gamma \overline{\Sigma}(\alpha) \Gamma^T,\\
				\label{eq:Psi block diagonal matrix}
				& = {\rm blkdiag}(\bar{E}_{1},\cdots,\bar{E}_{N}),
			\end{align}
		\end{subequations}
		where $L$ is defined in \eqref{eq:Loutput}, $\Gamma \in \mathbb{R}^{NM\times NM}$ is a lower unitriangular matrix with $|\Gamma|=1$ and for $j,i=1,\cdots,NM$, the $(j,i)$th element of $\Gamma$ is 
		\begin{align}\label{eq:def of Gamma}
			[\Gamma]_{ji} = \left\{ 
			\begin{aligned}
				0, \quad & j<i , \\
				1, \quad & j=i ,\\
				b_{j,i}, \quad &  j>i,
			\end{aligned}
			\right.
		\end{align}
	and ${\rm blkdiag}(\bar{E}_{1},\cdots,\bar{E}_{N})$ is a block diagonal matrix with 
		$\bar{E}_{1},\cdots,\bar{E}_{N}$,  defined in \eqref{eq:def of E_bar_j}, on the main diagonals.
	\end{Proposition} 
    
    By Proposition \ref{prop:diagonal structure of Psi}, the cost function of the MLM method \eqref{eq:loglikelihood} can be calculated as shown in the proposition below. 
    
    \begin{Proposition}\label{prop:computation of MLM cost function}
    	The cost function of the MLM method \eqref{eq:loglikelihood} can be computed by using
    	\begin{subequations}
    		\label{eq:relation3}
    	\begin{align}
    		\label{eq:logSigma}
    		&\log |\Sigma(\alpha)| = \sum^{N}_{j=1}\log|\bar{E}_j|,\\
    		\label{eq:YinvSigmaY}
    		&Y^T\Sigma^{-1}(\alpha)Y = \sum_{j=1}^N \bar{e}_j^T\bar{E}_j^{-1}\bar{e}_j.
    	\end{align}
    \end{subequations}
    \end{Proposition}

   Then by Propositions \ref{prop:diagonal structure of Psi} and \ref{prop:computation of MLM cost function}, the cost functions of the GCV and SURE methods can be calculated as shown in the following proposition.
   \begin{Proposition}\label{prop:computation of GCV and SURE cost function}
   	The cost functions of the GCV method \eqref{eq:GCV} and the SURE method \eqref{eq:SURE} can be computed by using
   	\begin{align}\label{eq:S and delta}
   		S =&\sigma^4 \sum_{j=1}^{N}\big[\bar{e}_j^T \bar{E}_j^{-1} (\bar{H} \bar{P}_{j|j-1} \bar{H}^T + I_M)
   		\bar{E}_j^{-1} \bar{e}_j
   		\nonumber\\&+ 2\bar{\zeta}_{j|j-1}^T \bar{H}\bar{E}_j^{-1} \bar{e}_j \big], \\
   		\delta =& MN - \sigma^2 \sum_{j=1}^N{\rm trace} \left[ \bar{E}_j^{-1}(\bar{H} \bar{P}_{j|j-1} \bar{H}^T + I_M)\right],\nonumber
   	\end{align}
    where $\bar{\zeta}_{j|j-1}$ and $\bar{P}_{j|j-1}$ can be computed recursively:
    \begin{itemize} 
    	\item for $j=1$, 
       $\bar{\zeta}_{1|0}=0\in\R^{Mr}$ and $\bar{P}_{1|0}=0\in\R^{Mr\times Mr}$;
    	\item for $j=2,\cdots,N$,
   \begin{align}
   	\label{eq:Nzeta}
   	\bar{\zeta}_{j|j-1} &= \bar{F}\bar{\zeta}_{j-1|j-2}+\bar{F}\bar{P}_{j-1|j-2}\bar{H}^T
   	\bar{E}_j^{-1}\bar{e}_j \nonumber\\& - \bar{F}\bar{\Sigma}_{j-1|j-2}\bar{H}^T\bar{E}_j^{-1}
   	(\bar{H}\bar{P}_{j|j-1}\bar{H}^T+I_M)\bar{E}_j^{-1}\bar{e}_j
   	\nonumber\\&- \bar{F}\bar{\Sigma}_{j-1|j-2}\bar{H}^T\bar{E}_j^{-1}\bar{H}\bar{\zeta}_{j-1|j-2},\\
   	\bar{P}_{j|j-1} &= \bar{F}\bar{P}_{j-1|j-2}\bar{F}^T - \bar{F}\bar{P}_{j-1|j-2}
   	\label{eq:NPk}
   	\bar{H}^T\bar{E}_j^{-1}\bar{H} \overline{\Sigma}_{j-1|j-2}\bar{F}^T 
   	\nonumber\\&- 
   	\bar{F}\overline{\Sigma}_{j-1|j-2}\bar{H}^T \bar{E}_j^{-1} \bar{H}\bar{P}_{j-1|j-2}\bar{F}^T
   	\nonumber\\ &-\bar{F}\overline{\Sigma}_{j-1|j-2}\bar{H}^T \bar{E}_j^{-1}
   	(\bar{H} \bar{P}_{j-1|j-2} \bar{H}^T+I_M)
   	\nonumber\\&\bar{E}_j^{-1}\bar{H}\overline{\Sigma}_{j-1|j-2}\bar{F}^T.
   \end{align}
\end{itemize}
   \end{Proposition}

	\subsection{Summary of the Implementation Algorithm and Its Computational Complexity Analysis} 
	
	The proposed implementation, as shown in Sections \ref{subsec:transformed SSM}-\ref{subsec:HypEst}, can be summarized in Algorithm \ref{alg:The Proposed Implementation} below.
	\begin{algorithm}
		\caption{The Proposed Implementation}
		\label{alg:The Proposed Implementation}
		\begin{algorithmic}
			\Require data $\{p_i,t_j,y_{i,j}\}_{i=1,j=1}^{M,N+N_T}$, kernels $k_t(t_j,t_{j'};\alpha_t)$, $k_s(p_i,p_{i'};\alpha_s)$
			\Ensure  $\hat{f}_{j|N}$ and $\overline{\Sigma}_{j|N}$ for $j = 1,\cdots,N+N_{T}$.  
			\Statex {\bf Step 1:} State-space model derivation
			\State \quad {Derive \eqref{eq:NewDTstate};}
			\Statex {\bf Step 2:} State-space model transformation
			\State \quad Calculate \eqref{eq:SVDofKs}, 
			\eqref{eq:F_bar H_bar} and \eqref{eq:Loutput};
			\Statex {\bf Step 3:} Hyper-parameter Estimation
			\Statex \quad  $\bullet$ Kalman filter 
			\State \quad \qquad Calculate  \eqref{eq:NewKalmanFilter};

			\Statex \quad \textbf{if} use the MLM method \eqref{eq:loglikelihood} \textbf{then}
			\State  \qquad Calculate \eqref{eq:relation3};
			\Statex \quad \textbf{end}
			\Statex \quad \textbf{if} use the GCV method \eqref{eq:GCV} \textbf{then}
			\State \qquad Calculate  \eqref{eq:Nzeta}, \eqref{eq:NPk},  \eqref{eq:eEe}, \eqref{eq:logEk} and \eqref{eq:recursivenewSdelta}; 
			\Statex \quad \textbf{end}
			\Statex \quad \textbf{if} use the SURE method \eqref{eq:SURE} \textbf{then}
			\State \qquad Calculate \eqref{eq:Nzeta}, \eqref{eq:NPk}, 
			 \eqref{eq:eEe}, \eqref{eq:logEk} and \eqref{eq:recursivenewSdelta};
			\Statex \quad \textbf{end}
			\Statex {\bf Step 4:} Function estimation and prediction

			\Statex \quad $\bullet$ Kalman smoother	for estimation
			\State \quad \qquad Calculate \eqref{eq:NewKalmanSmoother};
			\Statex \quad $\bullet$ Kalman predictor for prediction		
			\State \quad \qquad Calculate \eqref{eq:NewKalmanPredictor};
		\end{algorithmic}
	\end{algorithm}
To analyze the computational complexity of Algorithm \ref{alg:The Proposed Implementation}, it should be noted that the dimension $r$ of the state-space model \eqref{eq:StrictlyProperStateSpaceRepresentationd} is determined by the temporal kernel \eqref{eq:dtkernel} and is irrespective of, and often much smaller than, $M$ and $N$, and thus in what follows, we ignore $r$ and moreover, let $N_T=N$ in the analysis for brevity.
	
	\begin{theorem} \label{thm:computational complexity}
		The proposed implementation, as shown in Algorithm \ref{alg:The Proposed Implementation}, has computational complexity $\mathcal{O}(M^3+NM^2)$. In particular, 
		
		\begin{itemize}
			\item the state-space model transformation \eqref{eq:SVDofKs}, 
			\eqref{eq:F_bar H_bar} and \eqref{eq:Loutput} has computational complexity  $\mathcal{O}(M^3+NM^2)$; 
			
			\item 	the Kalman filter \eqref{eq:NewKalmanFilter} has computational complexity  $\mathcal{O}(NM)$;
			
			\item  the evaluation of the cost functions of the MLM method \eqref{eq:loglikelihood}, the GCV method \eqref{eq:GCV} and the SURE method \eqref{eq:SURE} has computational complexity $\mathcal{O}(NM)$;

			\item the Kalman smoother \eqref{eq:NewKalmanSmoother} and Kalman predictor \eqref{eq:NewKalmanPredictor} have computational complexity  $\mathcal{O}(NM^2)$.
			
		\end{itemize}

	\end{theorem}

\begin{remark}\label{rmk:optimization}
For spatial-temporal data with large $N$ and moderately large $M$, to reduce the computational complexity of the MLM method \eqref{eq:loglikelihood}, the GCV method \eqref{eq:GCV} and the SURE method \eqref{eq:SURE}, it is suggested to use derivative-free optimization algorithms or algorithms that only require numerical gradient,  approximated by finite difference of the cost function of the optimization problems involved. With such optimization algorithms, solving the MLM method \eqref{eq:loglikelihood}, the GCV method \eqref{eq:GCV} and the SURE method \eqref{eq:SURE} only involves the state-space model transformation, the Kalman filter and the evaluation of the cost function of the optimization problems and thus has computational complexity $\mathcal{O}(M^3+NM^2)$.

\end{remark}

	\section{Applications}\label{sec:Case}
	
	In this section, we illustrate the proposed implementation over applications in weather data prediction and spatially-distributed system identification.
	

	\subsection{Computing Platform}\label{sec:4Platform}
\begin{figure}[htbp]
		\centering
		\includegraphics[width=0.5\hsize]{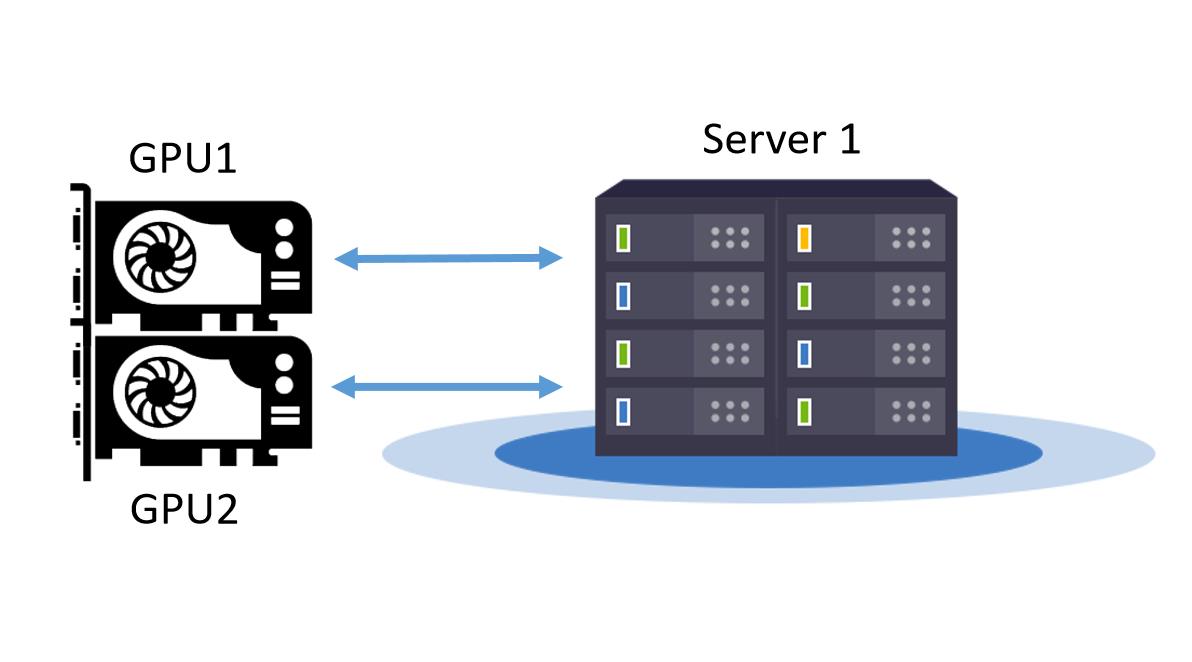}
		\caption{Computing platform}
		\label{fig:topology}
	\end{figure}

	Firstly, we introduce our computing platform  in Fig. \ref{fig:topology}, which consists of 1 server and 2 GPUs:
	\begin{itemize}
		\item Server 1: Intel(R) Xeon(R)  Platinum 8168 2.7GHz CPU$\times$2  (48 cores), 64GB$\times$24=1.48TB RAM,
		\item GPU: NVIDIA V100 $\times$2, 16GB RAM.
	\end{itemize}
	
	It is worth to note that many computations in the proposed implementation can be parallelized. For example, the creation of the spatial kernel matrix $K_s(\alpha_s)$, the SVD of $K_s(\alpha_s)$, the output transformation in \eqref{eq:lambdaT yj}, the computation of \eqref{eq:NfkN1} and \eqref{eq:hatf2}. Then by using the parallel computing structure of the computing platform and the parallel computing toolbox in MATLAB, the proposed implementation can be made more efficient. 
	
%


   \subsection{Weather Data Prediction}\label{sec:weather data}
   
   \subsubsection{Weather Data Sets}\label{subsubsec: weather data sets}
  
  We consider the following two weather data sets. 
  
  \begin{enumerate}[1)]
  	\item {\bf Colorado Precipitation Data}:
  	This data set has been tested in e.g., \cite{KCYZ19,TCCPS20}, contains monthly precipitation data between $1895$ and $1997$ from 367 weather stations in Colorado, USA\footnote{\tiny https://www.image.ucar.edu/Data/US.monthly.met/CO.shtml.}. The data set contains in total 1236 time instants and 367 locations (stations) located in a rectangular longitude/latitude region [109.5$^{\circ}$W, 101$^{\circ}$W]$\times$[36.5$^{\circ}$N, 41.5$^{\circ}$N]. 
  	%
  	We treat the data in $1895-1995$ as the training data, and the data in $1996-1997$ as the test data, that is, we have $t_j=jT_s$, $j=1,\cdots, N+N_T$ with $T_s=1$ month, $N=1212$, $N_T=24$ and $M=367$. This data set contains in total $453,612$ data points.
  	
  	\item {\bf GHCN Temperature Data}: 
  	This data set is obtained from the Global Historical Climatology Network (GHCN), and contains daily average temperatures collected from over ten thousands weather stations over the world \cite{GHCNdata}. We first choose 4000 stations with most complete data records from the $301$th day of 1999 to the $300$th day of 2018. Then we take out those stations with daily average temperature over 80$^{\circ}$C or under $-80^{\circ}$C and there are 3955 locations left.
  	The data set contains in total $6940$ times instants and $3955$ locations (stations) and we treat the data in the former 18 years as the training data, and the data in the last year as the test data, that is, we have $t_j=jT_s$, $j=1,\cdots, N+N_T$ with $T_s=1$ day,  $N=6575$, $N_T=365$ and $M=3955$. This data consist of more than 27 $million$ data points and is much larger than the Colorado precipitation data.
  	
  	
  \end{enumerate}
  
 \subsubsection{Kernel Design} \label{sec:kerneldesign}
 
In this section, we design a kernel for both the Colorado precipitation data and GHCN temperature data. The kernel design problem here is tricky, because the designed kernel should on the one hand incorporate the prior knowledge on the underlying spatial-temporal function to be estimated and on the other hand has state-space model realizations.

We first consider the spatial kernel design. Since the precipitation and the temperature are diffusion processes, the spatial prior knowledge is that for two locations, the closer the two locations, the larger the correlation between their weather data, and thus the squared exponential (SE) kernel is often adopted, e.g., \cite{WR06,TCCPS20},
\begin{align} \label{eq:spatialkernel}
	k_{\text{SE}}(p_{i},p_{i'};\alpha_{se})= \exp\left(-||p_{i}-p_{i'}||^2_2/\alpha_{se}\right),
\end{align}  
where $\alpha_{se}>0$, and  
	for the Colorado precipitation data, $p_{i}\in\R^2$ and its components are the longitude and latitude of the location, respectively, and for the GHCN temperature data, $p_{i}\in\R^{3}$ and its components are the earth-centered earth-fixed (ECEF) coordinates of the locations considered and the units are in 10 kilometers. 

Then we consider the temporal kernel design. To capture the periodicity of the weather data, an intuitive way is to use the periodic kernel, e.g., \cite{WR06}
 	\begin{align}\label{eq:PeriodicKernel}
 		k_{\text{per}}(\tau;\delta_{t},c_{t}) = \delta_t \exp \left\{ -2 c_t \left[\sin(\pi \texttt{f} \tau) \right]^2 \right\},
 	\end{align}
 	where $c_{t}>0$, $\delta_t>0$ are the hyper-parameters, and $\texttt{f}\in\R $ is the period of the weather data. However, the periodic kernel \eqref{eq:PeriodicKernel} does not have a proper PSD and thus has no state-space model realization. To overcome this difficulty, we first consider the Taylor expansion of $\exp(x)$ at $x=0$ to the second-order, then replace $x$ by $-2 c_t [\sin(\pi \texttt{f} \tau )]^2 $ in the expansion, and finally, 
 	multiply it by an exponential kernel $k_{\text{EXP}}(\tau) = e^{-{|\tau|}/{\sigma_t}}$ and obtain the following positive definite kernel  
 	\begin{align}
 		&k_{\text{TE2}}(\tau;\delta_{t},c_{t})k_{\text{EXP}}(\tau)\nonumber\\ \label{eq:Exp2tp}
 		&=\delta_t \left[ (1-c_t+\frac{3}{4}c_t^2)+(c_t-c_t^2)\cos(2\pi \texttt{f}|\tau|)   \right.\\\nonumber
 		&\left. 
 		+\frac{c_t^2}{4}\cos(4\pi \texttt{f}|\tau|) \right]\exp\left(-\frac{|\tau|}{\sigma_t}\right), 
 	\end{align} 
 	where $\delta_t>0$ and $c_t\in(0,1)$ are the hyper-parameters, $c_t\in(0,1)$ is imposed to  guarantee that \eqref{eq:Exp2tp} is positive semidefinite. The derivation of the state-space model of \eqref{eq:Exp2tp} is included in Appendix \ref{subsec:state-space realization of kernel^tpe}. Here, it should be noted that both $\texttt{f}$ and $\sigma_t$ are not hyper-parameters: \texttt{f} is chosen to be $\texttt{f} =1/12$ for the Colorado precipitation data and $\texttt{f}=1/365.3$ for the GHCN temperature data due to the periodicity of the data, and $\sigma_t$ is chosen to be $\sigma_t = 5\times 10^3$ such that the exponential kernel $k_{\text{EXP}}(\tau) = e^{-{|\tau|}/{\sigma_t}}$ has a negligible effect.  
Moreover, to describe the slight temporal variation of the data, we further include a Mat\'ern kernel, e.g., \cite{WR06}, i.e., 
 		\begin{align}\label{eq:matern kernel}
 			k_{\text{Matern}}(\tau; h_{t},\theta_{t})=h_t\left(1+\frac{\sqrt{3}|\tau|}{\theta_t}\right) \exp\left(-\frac{\sqrt{3}|\tau|}{\theta_t}\right),
 		\end{align}
 		where $h_{t},\theta_t >0$ are the hyper-parameters of \eqref{eq:matern kernel}. Then we can obtain the following temporal kernel
 	\begin{align}\label{eq:temporalkernel3}
 		k_t(\tau;\alpha_{tm})=k_{\text{TE2}}(\tau;\delta_{t},c_{t})k_{\text{EXP}}(\tau) + k_{\text{Matern}}(\tau; h_{t},\theta_{t}),
 	\end{align}
 	where $\alpha_{tm}=[\delta_{t}, c_t,  h_t, \theta_t]^T$ with  
 	\begin{align}\label{eq:smallvariation}
 		0.01\delta_t\leq h_t\leq 0.1\delta_t,
 	\end{align}
 	which is enforced to guarantee that the Mat\'ern kernel \eqref{eq:matern kernel} describes the slight temporal variation of the data.
 	
 	\begin{remark}\label{rmk: temporal kernel}
 	Beside the spatial prior knowledge considered above, it is interesting to note that the spatial prior knowledge considered in  \cite{Zorzi20} is that the edges in the graphical model are sparse, where the graphical model is due to the existence of a number of modules with a graphical structure, and that each module has a number of nodes sharing the same graphical structure, and thus, a sparsity inducing kernel/regularization was designed accordingly. 
 		It is also interesting to mention that the following kernel 
 		\begin{align}\label{eq:periodicallydecaying}
 			k_{\text{PD}}(\tau;\delta_{t},\sigma_t) =\delta_{t} \cos(2\pi \texttt{f}|\tau|)e^{-\frac{|\tau|}{\sigma_t}},
 		\end{align}
 		where $\delta_{t},\sigma_t >0$ are hyper-parameters and $\texttt{f}=1/12$, is chosen in \cite{TCCPS20,KCYZ19} as the temporal kernel. 

 	\end{remark}

   	\subsubsection{Hyper-parameter Estimation and Function Prediction}\label{subsubsec:hyper-parameter est} 
   
   For the two data sets and designed kernels, we use the MLM, GCV and SURE methods, as shown in Section \ref{subsec:HypEst}, to estimate the hyper-parameter $\alpha = [\alpha_t^T,\alpha_s^T ]^T$. 
   Moreover, for the MLM method, the noise variance $\sigma^2$ is treated as an additional hyper-parameter, i.e., $\alpha=[\alpha_{t}^T, \alpha_{s}^{T}, \sigma^2]^T$, and its estimate is then used for the SURE and GCV methods. With the estimated hyper-parameter, we can further run the Kalman filter, smoother and predictor in Section \ref{subsec:KSKP} to compute the function prediction $\hat{f}_T = [\hat{f}_{N+1|N},\cdots,\hat{f}_{N+N_T|N}]$.  
   
   The function $\texttt{fmincon}$ in Matlab, using the interior-point algorithm with numerical gradient approximated by finite difference of the cost function,  is applied to solve \eqref{eq:loglikelihood}, \eqref{eq:GCV} or \eqref{eq:SURE}. Since the selection of initial points is significant for the search of ``good'' local minima, the following way is used to find  a ``good'' local minimum: 
   	\begin{enumerate}
   		\item [1)] for each component of the hyper-parameter, we select a set of initial points and thus obtain a grid of initial points of the hyper-parameter;
   		\item [2)] calculate the cost functions over the grid of initial points;
   		\item [3)] select 5 initial points corresponding to the smallest 5 values of the cost function;
   		\item [4)] use the function $\texttt{fmincon}$ with selected 5 initial points to solve the optimization problem involved in the hyper-parameter estimation, respectively;
   		\item [5)] choose the optimal solution with the smallest value of the cost function as the  optimal hyper-parameter estimate.
   	\end{enumerate}

   To assess how good the prediction $\hat{f}_T $ is at $jT_s$ for $j = N+1,\cdots,N+N_T$, we use the measure of fit, e.g., \cite{Ljung:00},
   \begin{align} \label{eq:def of fitj}
   	\text{fit}_{j} = 100\left( 1 - \frac{||\hat{f}_{j|N}-y_j||_2}{||y_j- \overline{y_j}||_2} \right), \overline{y_j}=\frac{1}{M}\sum_{i=1}^{M}y_{i,j}.
   \end{align}
   The maximum of $\text{fit}_{j}$ is 100, meaning a perfect match between $\hat{f}_{j|N}$ and $y_j$. The average prediction fit over the test data set is defined as
   \begin{align}\label{eq:def of average fit}
   	\overline{\text{fit}} =\frac{1}{N_{T}}\sum_{j=N+1}^{N+N_{T}}\text{fit}_{j}.
   \end{align}
   
\vspace*{-4mm}

   \subsubsection{Filling the Missing Data}
   
   The Colorado precipitation and GHCN temperature data contain  $58.39\%$  and $3.3\%$ missing data, respectively, and we need to fill the missing data before running simulations. To this goal, we first split the spatial-temporal data into $M$ temporal data sets $\{p_{1},t_{j},y_{1,j}\}_{j=1}^{N+N_{T}},\cdots,\{p_{M},t_{j},y_{M,j}\}_{j=1}^{N+N_{T}}$ according to the $M$ locations. 
   For each temporal data set, the temporal kernel \eqref{eq:periodicallydecaying} or \eqref{eq:temporalkernel3} is applied, respectively.
   Then for each $i=1,\cdots,M$, we use the MLM method to estimate the corresponding hyper-parameter and in particular, if $y_{i,j}$ is missing for some $j=1,\cdots,N+N_{T}$, then no measurement update is needed, i.e.,  
   \eqref{eq:Nxkk} and \eqref{eq:NSigmakk} should be replaced by 
   \begin{align}
   	\hat{x}_{j|j} = \hat{x}_{j|j-1} ,\quad \overline{\Sigma}_{j|j} = \overline{\Sigma}_{j|j-1},
   \end{align}
   respectively, e.g., \cite{AndersonM:79}. Finally, with the obtained hyper-parameter estimate, the Kalman smoother \eqref{eq:NewKalmanSmoother} is used to fill the missing data.
   
   \begin{remark}
   	The above treatment of the missing data implicitly assumes that for $i,i'=1,\cdots,M$ and $j,j'=1,\cdots,N+N_{T}$, if $i\neq i'$, $y_{i,j}$ and $y_{i',j'}$ are independent. The treatment in \cite{TCCPS20} does not rely on this assumption and thus is more general but with the price of higher computational complexity.  
   \end{remark}

\subsubsection{Illustration of Computational Efficiency}

Firstly, we consider the Colorado precipitation data and choose \eqref{eq:spatialkernel} as the spatial kernel and \eqref{eq:periodicallydecaying} as the temporal kernel, and then we evaluate  the cost functions of the MLM method \eqref{eq:loglikelihood}, GCV method \eqref{eq:GCV}, and SURE method \eqref{eq:SURE} for 10 times. The average computing time of the cost functions for the proposed implementation and the one in \cite{TCCPS20,KCYZ19} are shown in the Table \ref{table:avgtimeObj}, which shows that, our proposed implementation is over 300 and 200 times faster than the one in \cite{TCCPS20,KCYZ19} for the MLM method, and GCV and SURE methods, respectively. 
\begin{table}[!htb]
	\centering
	\caption{The average computing time (in second) of the cost functions of the MLM method \eqref{eq:loglikelihood}, GCV method \eqref{eq:GCV} and SURE method \eqref{eq:SURE} for the Colorado precipitation data.}
	\label{table:avgtimeObj}
	\begin{tabular}{ccc}
		\hline
		Implementation & Proposed & in \cite{TCCPS20,KCYZ19}  \\
		\hline
		MLM method& 0.6485 & 197.3668 \\ 
		GCV method & 1.5983 & 332.9270  \\
		SURE method & 1.5904 & 331.9392\\ 	 
		\hline 
	\end{tabular}
\end{table} 
\vspace*{-6mm}


\begin{figure}[H]
	\centering
	\includegraphics[width=0.4\textwidth,height=0.25\textwidth]{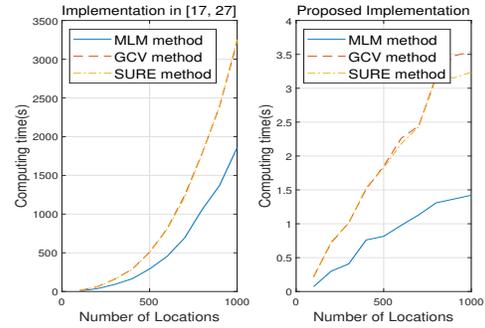}
	\caption{The average computation time (in second) of the cost functions of the MLM method \eqref{eq:loglikelihood}, GCV method \eqref{eq:GCV} and SURE method \eqref{eq:SURE} for the GHCN temperature data with $N=800$ and $M=100,200,\cdots,1000$, respectively, where our proposed implementation and the one in \cite{TCCPS20,KCYZ19} are shown on the right and left panels, respectively. 
	}
	\label{fig:efficiency compare} 
\end{figure}
\vspace*{-5mm}
Secondly, we consider the GHCN temperature data but only use part of it, because the implementation in \cite{TCCPS20,KCYZ19} is too expensive to be applied to the full data. In particular, we only use the first 800 time instants and 1000 locations, i.e.,  
$\{p_i,t_j,y_{i,j}\}_{i=1,j=1}^{1000,800}$. Then we choose \eqref{eq:spatialkernel} and \eqref{eq:temporalkernel3} as the spatial kernel and temporal kernel, respectively, and  evaluate the cost functions of the MLM method \eqref{eq:loglikelihood}, GCV method \eqref{eq:GCV}, and SURE method \eqref{eq:SURE} for 10 times with $N=800$ and $M=100,\cdots,1000$, respectively. The average computing time of the cost functions for the proposed implementation and the one in \cite{TCCPS20,KCYZ19} are shown Fig. \ref{fig:efficiency compare}, which shows that, our proposed implementations is more efficient than the one in \cite{TCCPS20,KCYZ19}, as the number of the locations increases. It is worth to mention that for the full GHCN temperature data, our proposed implementation has the average computing time 30.2, 68.2 and 67.8 seconds, for the cost functions of the MLM, GCV, and SURE methods, respectively.

\subsubsection{Illustration of Prediction Performance}\label{subsubsec:illustration of performance}

\begin{figure*}[thpb]
	\centering
	\begin{subfigure}[b]{0.3\textwidth}
		\centering
		\includegraphics[width=\textwidth,height=0.6\textwidth]{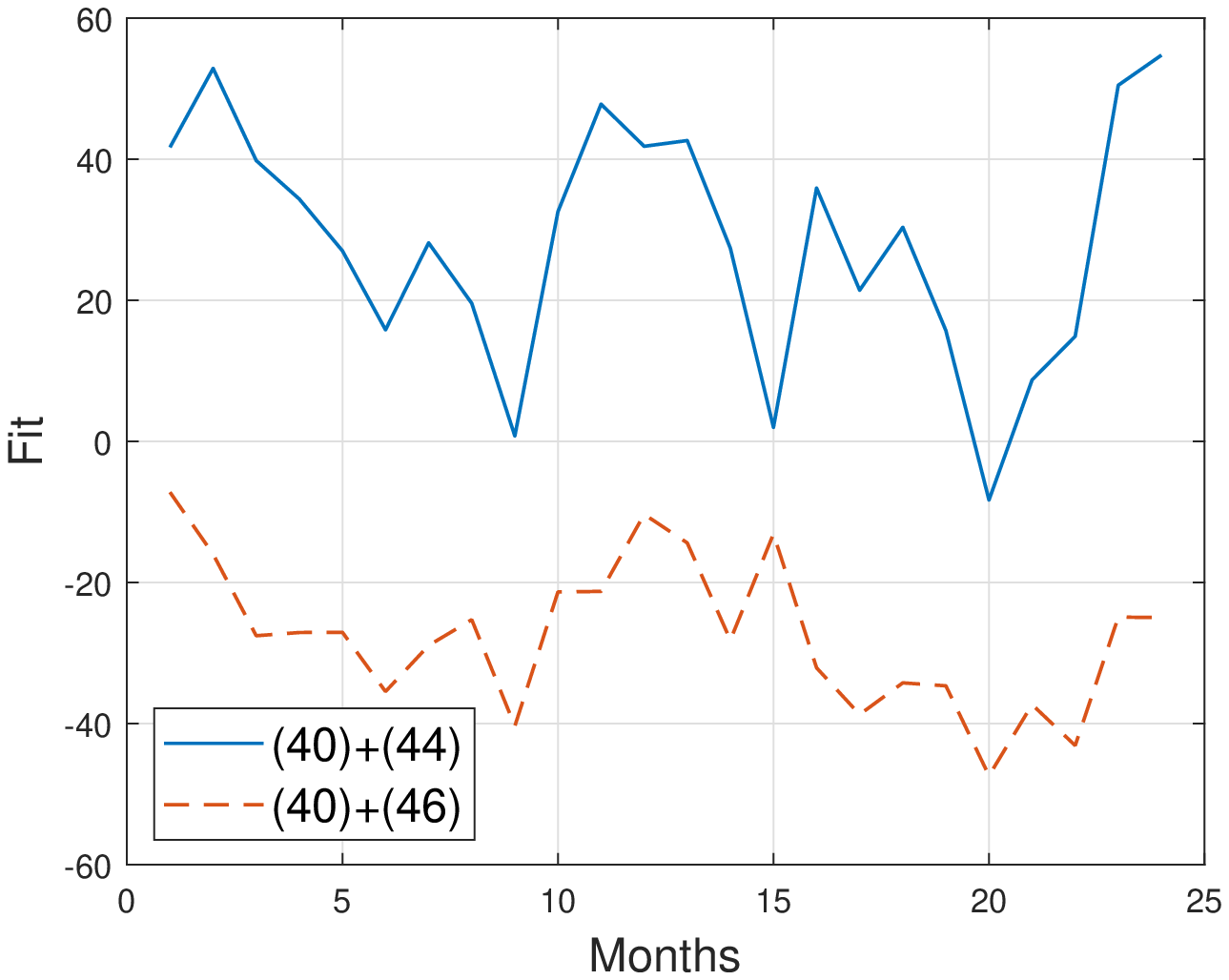}
		\caption{}
		\label{fig:ColoradoPredictionfitComparisonML}
	\end{subfigure}
	\begin{subfigure}[b]{0.3\textwidth}
		\centering
		\includegraphics[width=\textwidth,height=0.6\textwidth]{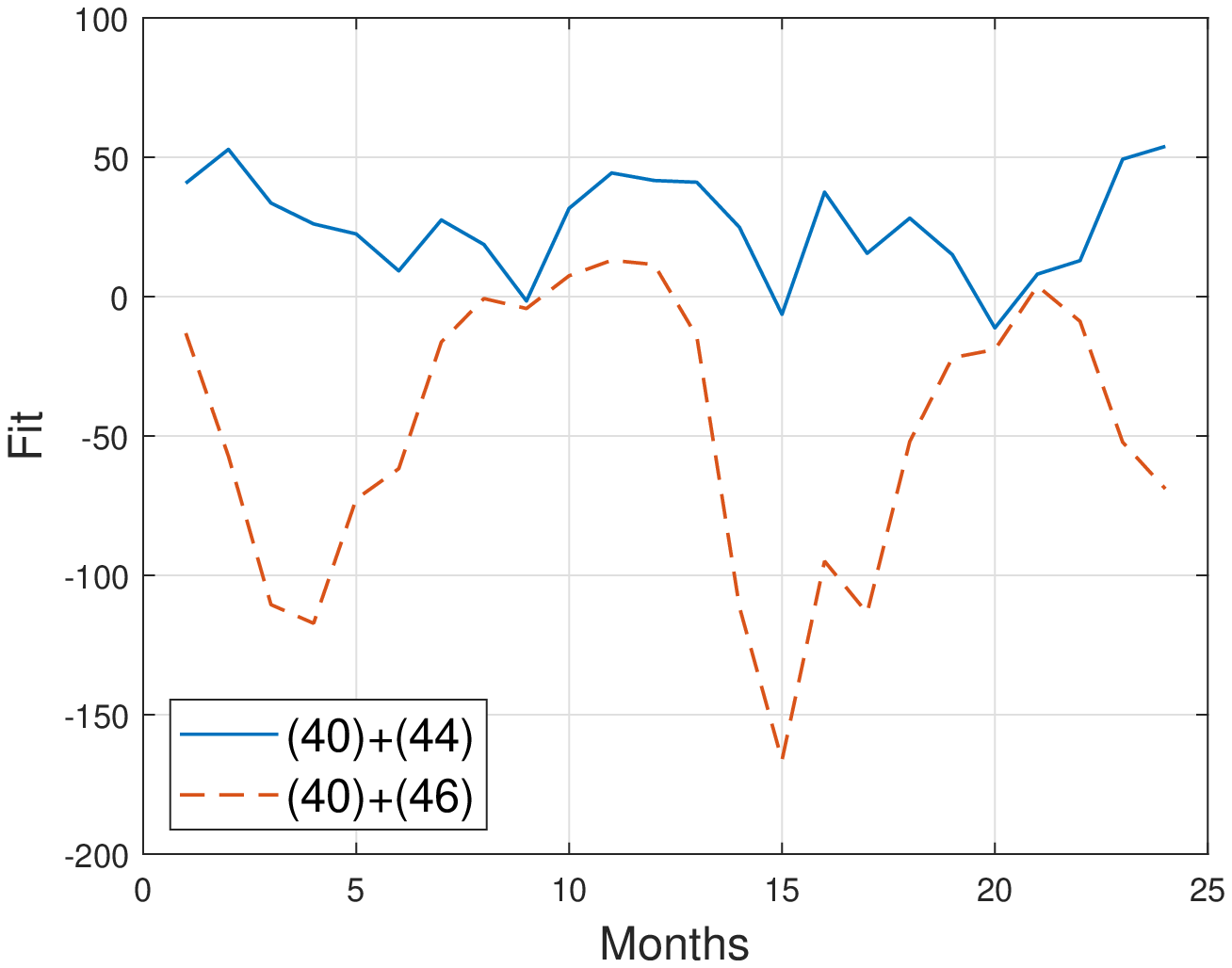}
		\caption{}
		\label{fig:ColoradoPredictionfitComparisonGCV}
	\end{subfigure}
	\begin{subfigure}[b]{0.3\textwidth}
		\centering
		\includegraphics[width=\textwidth,height=0.6\textwidth]{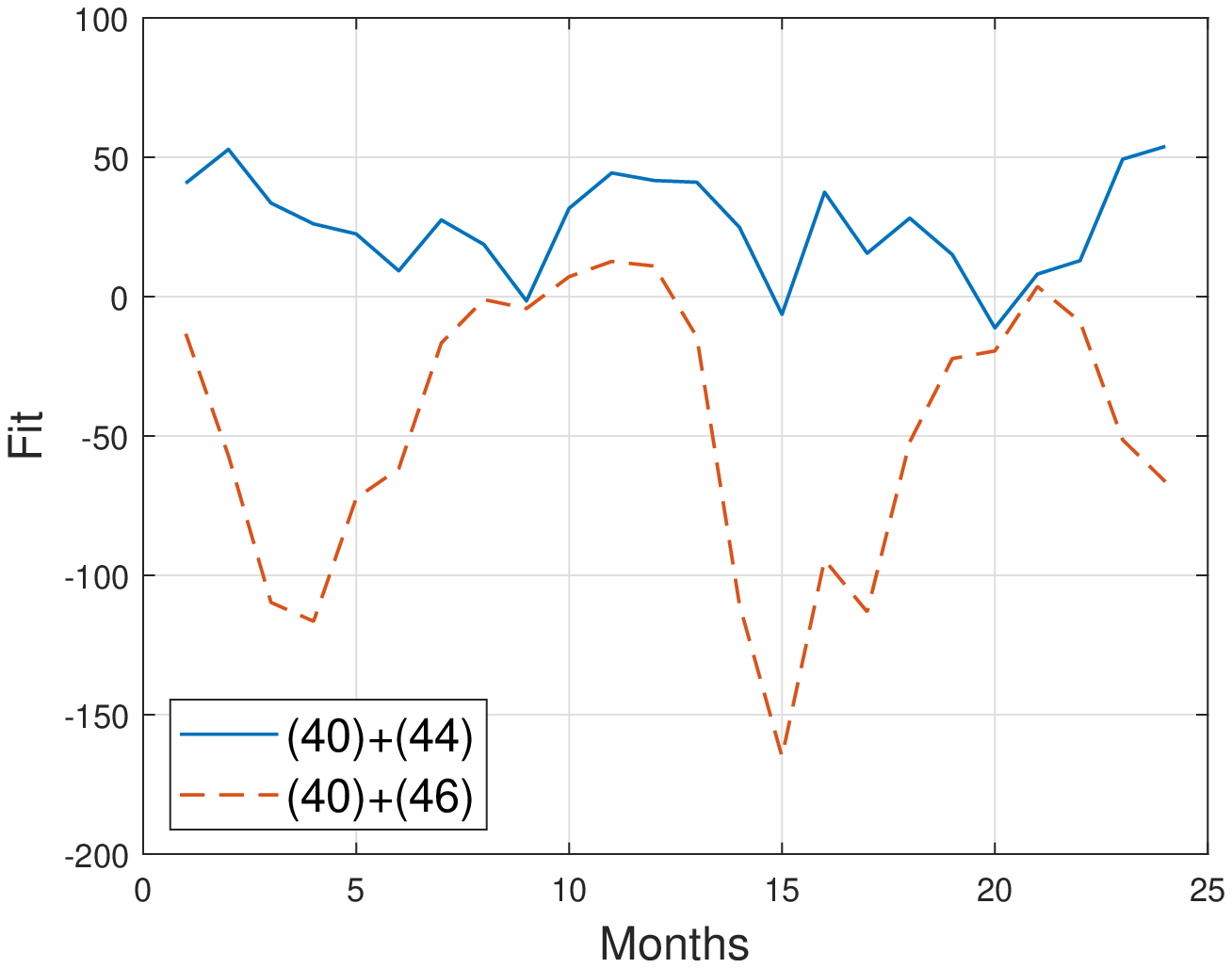}
		\caption{}
		\label{fig:ColoradoPredictionfitComparisonSURE}
	\end{subfigure}
	\caption{Profile: Monthly prediction fits \eqref{eq:def of fitj} for the Colorado precipitation data using two kernel combinations: \eqref{eq:spatialkernel}+\eqref{eq:temporalkernel3} and  \eqref{eq:spatialkernel}+\eqref{eq:periodicallydecaying}, and three hyper-parameter estimation methods, respectively. Panel (a): Monthly prediction fits using the MLM method. Panel (b): Monthly prediction fits using the GCV method. Panel (c): Monthly prediction fits using the SURE method. }
	
	\label{fig:ColoradoFitML}
\end{figure*}

\begin{figure*}[thpb] 
	\centering
	%
	\begin{subfigure}[b]{0.3\textwidth}
		\centering
		\includegraphics[width=\textwidth, height=0.6\textwidth]{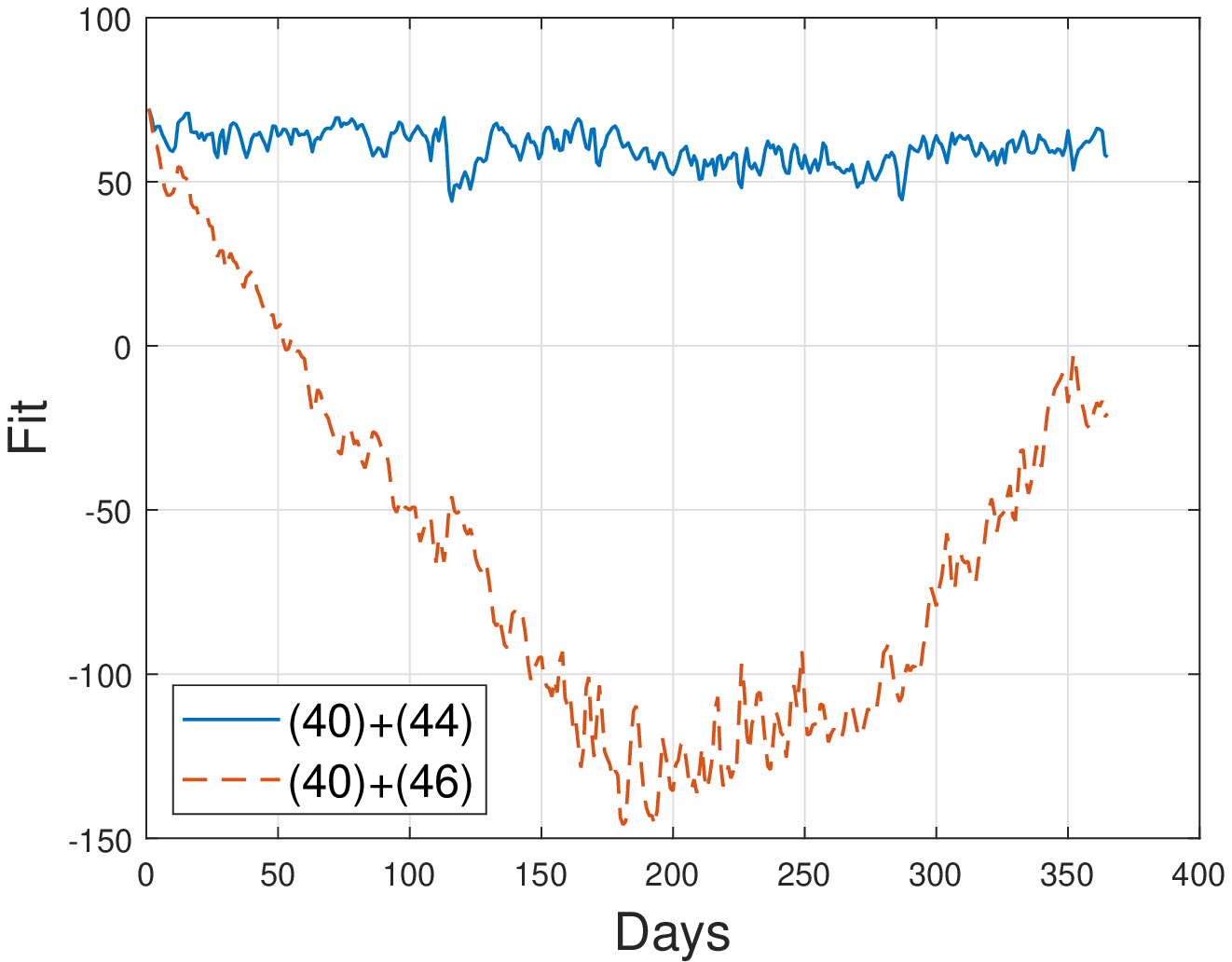}
		\caption{}
		\label{fig:ML3Kernels3delayGpe2M}
	\end{subfigure}
	\begin{subfigure}[b]{0.3\textwidth}
		\centering
		\includegraphics[width=\textwidth, height=0.6\textwidth]{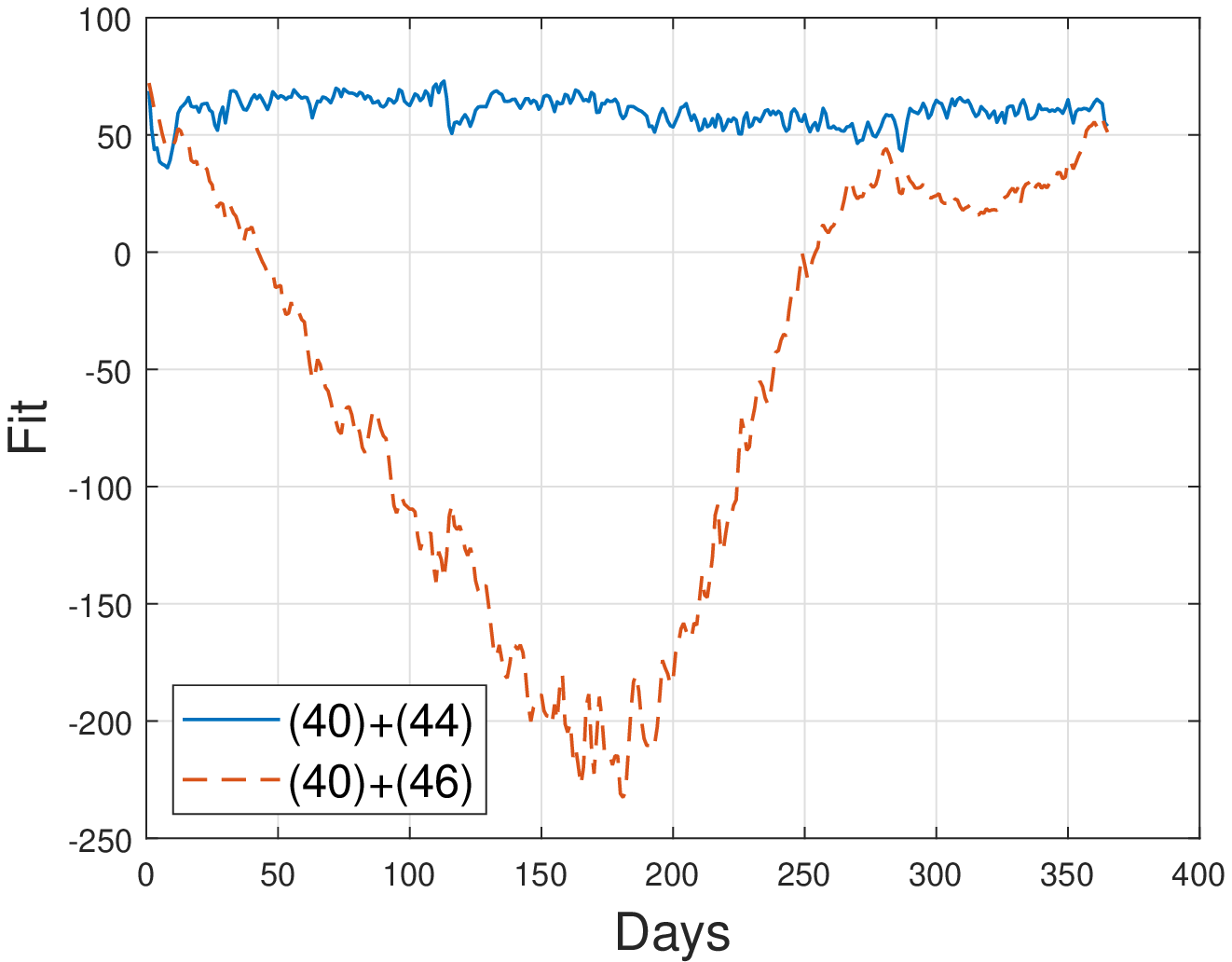}
		\caption{}
		\label{fig:GCV3Kernels3delayGpe2MS}
	\end{subfigure}
	\begin{subfigure}[b]{0.3\textwidth}
		\centering
		\includegraphics[width=\textwidth, height=0.6\textwidth]{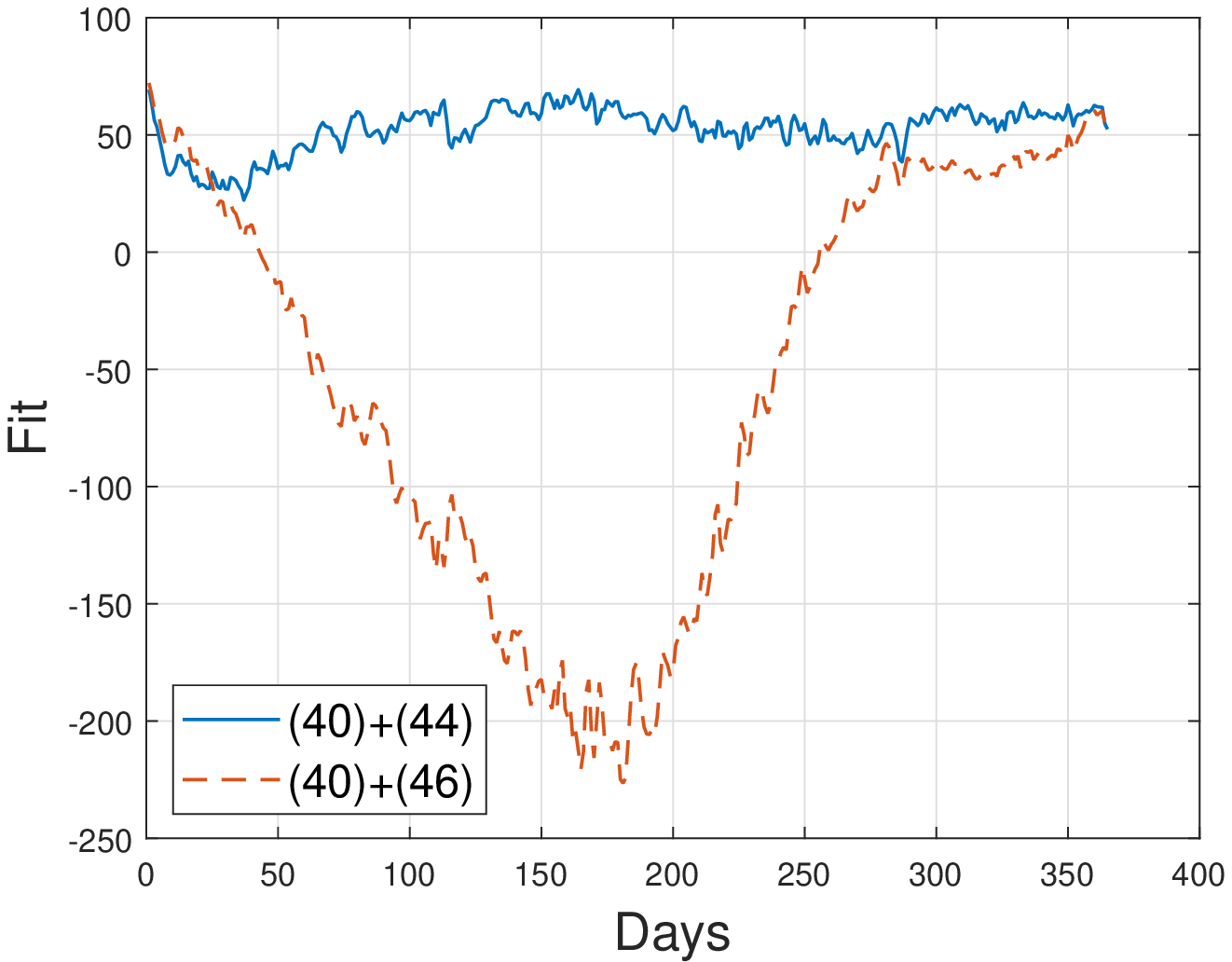}
		\caption{}
		\label{fig:SURE3Kernels3delayGpe2MS}
	\end{subfigure}
	\caption{Profile: Daily prediction fits \eqref{eq:def of fitj} for the GHCN temperature data using two kernel combinations: \eqref{eq:spatialkernel}+\eqref{eq:temporalkernel3} and  \eqref{eq:spatialkernel}+\eqref{eq:periodicallydecaying}, and three hyper-parameter estimation methods, respectively. Panels (a): Daily prediction fits using the MLM method. Panel (b): Daily prediction fits using the GCV method. Panel (c): Daily prediction fits using the SURE method.}
	\label{fig:MLGCVSUREKernelsdelay}
\end{figure*}
\begin{table}[b]
	\centering
	\caption{The average prediction fits \eqref{eq:def of average fit} for the Colorado precipitation data using two kernel combinations, where the hyper-parameters are estimated by MLM, GCV and SURE methods, respectively. The values in parentheses are the corresponding smallest prediction fits.}
	\label{tab:precipitation}
	\begin{tabular}{lll}

		\hline

		Kernel 	& \eqref{eq:spatialkernel}+\eqref{eq:temporalkernel3} & \eqref{eq:spatialkernel}+\eqref{eq:periodicallydecaying} \\

		\hline

		MLM  & 28.25 (-8.31)  & -27.54 (-47.39)\\  

		GCV  & 25.66 (-11.24) & -47.56 (-166.35)\\ 

		SURE & 25.66 (-11.24) & -47.38 (-165.38) \\ 

		\hline 

	\end{tabular}
\end{table}

\begin{table*}[thtb]\scriptsize
	\centering
	\caption{The hyper-parameter and the corresponding optimal values of the cost functions for the Colorado precipitation data using two kernel combinations: \eqref{eq:spatialkernel}+\eqref{eq:temporalkernel3} and  \eqref{eq:spatialkernel}+\eqref{eq:periodicallydecaying}, and three hyper-parameter estimation methods, respectively. }
	\label{tab:Hyper-parameter of Colorado}
	\begin{tabular}{rcccccccc}

		\hline

		Method, Kernels & Optimal cost function value & $\delta_t$ & $\sigma_t$ & $\sigma^2$ & $\alpha_{se}$ & $c_t$ &$h_t$ &$\theta_t$  \\

		\hline

		MLM, \eqref{eq:spatialkernel}+\eqref{eq:temporalkernel3} & 1.8876e+06

		& 4.6377e+03 &5000 & 173.0302 & 0.1303 & 0.3834 & 109.4592 & 0.0663  \\ 

		MLM, \eqref{eq:spatialkernel}+\eqref{eq:periodicallydecaying} & 2.0311e+06 & 361.7502 & 2.2946 & 501.5663 & 8.0899 & $-$ & $-$ & $-$ \\ 




		GCV, \eqref{eq:spatialkernel}+\eqref{eq:temporalkernel3} & 242.3078

		& 9.3410e+03 &5000 & 173.0302 & 0.1373 & 0.5345 & 93.4106 &  0.1136 \\ 

		GCV, \eqref{eq:spatialkernel}+\eqref{eq:periodicallydecaying} & 508.7996 & 9.0237e+08 & 1.1407e+07 & 501.5655 & 2.5726 & $-$ & $-$ & $-$   \\


		SURE, \eqref{eq:spatialkernel}+\eqref{eq:temporalkernel3} & 9.8970e+07

		& 9.2773e+03 &5000& 173.0302 & 0.1376 & 0.5346 & 92.7727 & 0.0562 \\

		SURE, \eqref{eq:spatialkernel}+\eqref{eq:periodicallydecaying} & 2.2675e+08 & 5.0370e+08 & 6.8254e+06 & 501.5655 & 2.8876 & $-$ & $-$ & $-$  \\

		\hline 

	\end{tabular}
\end{table*}
\begin{table}[htb]
	\centering
	\caption{The average prediction fits \eqref{eq:def of average fit} for the GHCN temperature date using two kernel combinations, where the hyper-parameters are estimated by MLM, GCV and SURE methods, respectively. The values in parentheses are the corresponding smallest prediction fits.}
	\label{tab:MLMGCVSURE}
	\begin{tabular}{lll}
		\hline
		Kernel 	& \eqref{eq:spatialkernel}+\eqref{eq:temporalkernel3} & \eqref{eq:spatialkernel}+\eqref{eq:periodicallydecaying}\\
		\hline
		MLM   & 60.34 (44.12) &  -62.85(-146.20)\\
		GCV   & 60.30 (35.97) &  -57.48(-232.58)\\ 
		SURE  & 52.40 (22.20) &  -53.57(-226.53)\\ 
		\hline 
	\end{tabular}
\end{table}
\begin{table*}[htb] \tiny   
	\centering
	\caption{The hyper-parameter and the corresponding optimal values of the cost functions for the GHCN temperature data using two kernel combinations:  \eqref{eq:spatialkernel}+\eqref{eq:temporalkernel3} and \eqref{eq:spatialkernel}+\eqref{eq:periodicallydecaying}, and three hyper-parameter estimation methods, respectively. }
	\label{tab:Hyper-parameter of GHCN}
	\begin{tabular}{rcccccccc}
		\hline
		Method, Kernels & Optimal cost function value & $\delta_t$ & $\sigma_t$ & $\sigma^2$ & $\alpha_{se}$ & $c_t$ &$h_t$ &$\theta_t$ \\
		\hline
		MLM, \eqref{eq:spatialkernel}+\eqref{eq:temporalkernel3} & 5.4912e+07 & 585.9242 & 5000 & 2.6159 & 984.0928 & 0.2867 & 5.8592 & 1.9000 \\
		MLM, \eqref{eq:spatialkernel}+\eqref{eq:periodicallydecaying} & 5.7639e+07  & 626.4846 & 235.5695 & 1.1125 & 91.8811 & - & - & -\\
		GCV, \eqref{eq:spatialkernel}+\eqref{eq:temporalkernel3} &  2.1015  & 1.8304e+04 & 5000 & 2.6159 & 19.6885 & 0.6481 & 1.7664e+03 & 8.3648  \\ 
		GCV, \eqref{eq:spatialkernel}+\eqref{eq:periodicallydecaying} & 2.4189 & 2.4023e+04 & 6.1204e+03 & 1.1125 & 47.8679 & - & - & -\\
		SURE, \eqref{eq:spatialkernel}+\eqref{eq:temporalkernel3}  & 7.6828e+07 & 5.7973e+03 & 5000 & 2.6159 & 345.0477 & 0.4735 & 579.7336 & 50.3385\\
		SURE, \eqref{eq:spatialkernel}+\eqref{eq:periodicallydecaying} & 4.4955e+07 & 1.1622e+04 & 4.8269e+03 & 1.1125 & 64.0244 & - & - & -\\
		\hline 
	\end{tabular}
\end{table*}
For the Colorado precipitation data, the prediction fits \eqref{eq:def of fitj}, the average prediction fits \eqref{eq:def of average fit}, and the optimal hyper-parameters using two kernel combinations: \eqref{eq:spatialkernel}+\eqref{eq:temporalkernel3} and  \eqref{eq:spatialkernel}+\eqref{eq:periodicallydecaying}, and three hyper-parameter estimation methods: MLM, GCV and SURE, are shown in Fig. \ref{fig:ColoradoFitML}, Tables \ref{tab:precipitation} and \ref{tab:Hyper-parameter of Colorado}, respectively. 
It is worth to stress that \eqref{eq:spatialkernel}+\eqref{eq:periodicallydecaying} was used in \cite{KCYZ19,TCCPS20}.

Fig. \ref{fig:ColoradoFitML} and Table \ref{tab:precipitation} show that for the same hyper-parameter estimation method, the prediction fits \eqref{eq:def of fitj} and the average prediction fits \eqref{eq:def of average fit} obtained by the kernel combination \eqref{eq:spatialkernel}+\eqref{eq:temporalkernel3} are all larger than those by \eqref{eq:spatialkernel}+\eqref{eq:periodicallydecaying}, indicating that the temporal kernel \eqref{eq:temporalkernel3} can better describe the Colorado precipitation data than  \eqref{eq:periodicallydecaying} used in \cite{TCCPS20,KCYZ19}. 
Table \ref{tab:precipitation} also shows that the kernel combination \eqref{eq:spatialkernel}+\eqref{eq:temporalkernel3} with hyper-parameters estimated by the MLM method gives the best average prediction fit $28.25$. One may wonder why this fit is not so good and the reason is perhaps due to that $58.39\%$ of the Colorado precipitation data are missing.    


For the GHCN temperature data, the prediction fits \eqref{eq:def of fitj}, the average prediction fits \eqref{eq:def of average fit}, and the optimal hyper-parameters using two kernel combinations: \eqref{eq:spatialkernel}+\eqref{eq:temporalkernel3} and  \eqref{eq:spatialkernel}+\eqref{eq:periodicallydecaying}, and three hyper-parameter estimation methods: MLM, GCV and SURE, are shown in Fig. \ref{fig:MLGCVSUREKernelsdelay}, Tables \ref{tab:MLMGCVSURE} and \ref{tab:Hyper-parameter of GHCN}, respectively.
It is worth to stress that \eqref{eq:spatialkernel}+\eqref{eq:periodicallydecaying} was used in \cite{KCYZ19,TCCPS20}. 


	Fig. \ref{fig:MLGCVSUREKernelsdelay} and Table \ref{tab:MLMGCVSURE} show that for the same hyper-parameter estimation method, the prediction fits \eqref{eq:def of fitj} and the average prediction fits \eqref{eq:def of average fit} obtained by the kernel combination \eqref{eq:spatialkernel}+\eqref{eq:temporalkernel3} are most of time larger than those by \eqref{eq:spatialkernel}+\eqref{eq:periodicallydecaying}, indicating that the temporal kernel \eqref{eq:temporalkernel3} can better describe the GHCN temperature data than  \eqref{eq:periodicallydecaying} used in \cite{TCCPS20,KCYZ19}.   
Table \ref{tab:MLMGCVSURE} also shows, among three hyper-parameter estimation methods, the MLM method gives the best average prediction fit $60.34$ and the corresponding smallest prediction fit $44.12$. Moreover, Fig. \ref{fig:MLGCVSUREKernelsdelay} also shows that the prediction fit of the GCV and SURE methods drop down quickly in the beginning and then go up again, while the MLM method can avoid such drop. 

\subsection{Spatially-distributed System Identification}\label{subsec:spatially distributed sysid}

In this section, we consider the identification of {\it spatially-distributed} system, e.g. \cite{QHJ18}, which is a class of distributed parameter systems. 

First, we recall from e.g., \cite{QHJ18}, that the subsystem at the $i$th location $p_i$ with $i=1,\cdots,M$ of a spatially-distributed system can be described by the following ARX model
\begin{align}\label{eq:ARX model}
\begin{aligned}
	\mathcal{A}_i(q_{p},q_{t})f(p_{i},t_{j})=&\mathcal{B}_i(q_{p},q_{t})u(p_{i},t_{j}),\\
	& t_j=jT_s, j=1,2,\cdots,N,
\end{aligned}
\end{align}
where $q_{p}$ and $q_{t}$ are the forward spatial and temporal shift operators, respectively, i.e., $q_{p}q_{t}^{-1}f(p_{i},t_{j})=f(p_{i+1},t_{j-1})$, $f(p_{i},t_{j})\in\R$ and $u(p_{i},t_{j})\in\R$ are the output and input at the $i$th location $p_i$ and $j$th time instant $t_j$, respectively, and
\begin{align}
	\mathcal{A}_i(q_{p},q_{t})=&1+\sum_{k_{i}=1}^{n_{a}}\sum_{k_{j}=0}^{M}a_i(p_{k_{j}},t_{k_{i}})q_{p}^{-k_{j}}q_{t}^{-k_{i}},\\
	\mathcal{B}_i(q_{p},q_{t})=&\sum_{l_{i}=1}^{n_{b}}\sum_{l_{j}=0}^{M}b_i(p_{l_{j}},t_{l_{i}})q_{p}^{-l_{j}}q_{t}^{-l_{i}},
\end{align}
with $a_i(p_{k_{j}},t_{k_{i}}),b_i(p_{l_{j}},t_{l_{i}})\in\R$, $n_{a},n_{b}\in\mathbb{N}$ and $p_0$ a null position. Then, we consider a special case of \eqref{eq:ARX model} with
\begin{align}\label{eq:ARX_special}
\begin{aligned}
	\mathcal{A}_i(q_{p},q_{t})&=1,	\\\mathcal{B}_i(q_{p},q_{t})&=\sum_{k=1}^{n_{b}}b_i(p_i,t_k)q_{t}^{-k}, u(p_{i},t_{j})=u(t_{j}),
\end{aligned}
\end{align} which is equivalent to assume that the subsystem at the $i$th location $p_i$ with $i=1,\cdots,M$, has a finite impulse response (FIR) model. The FIR parameters $b_i(p_i,t_k)$, $k=1,\cdots,n_b$ only depend on $p_i$ and moreover, assumed to be smooth functions of $p_i$. In this case, the output $f(p_{i},t_{j})$ takes the form of 
\begin{align}
f(p_{i},t_{j})= \sum_{k=1}^{n_{b}}b_i(p_i,t_k) u(t_{j-k}),
\end{align} which plays the role as the spatial-temporal function $f(p_{i},t_{j})$ in \eqref{eq:NoisyObservation}.


In what follows, we study the identification of spatially-distributed system \eqref{eq:ARX model} with \eqref{eq:ARX_special}, i.e., the estimation of the FIRs $\{b(p_{i},t_{j})\}_{i=1,j=1}^{M,n_b}$ of $M$ spatially-distributed subsystems as well as possible based on the training data $\{y_{i,j},u(t_{j})\}_{i=1,j=1}^{M,N}$ by using the Gaussian process regression approach in this paper. For comparison, we also consider the estimation of the FIR $\{b(p_{i},t_{j})\}_{j=1}^{n_b}$ of the $i$th subsystem based on  $\{y_{i,j},u(t_{j})\}_{j=1}^{N}$ separately by neglecting the spatial interconnections between $M$ subsystems and by using the approach in \cite{COL12}. These two approaches are denoted by  the ``spatial-temporal'' and ``temporal'' approaches in the following, respectively.

\subsubsection{Test Spatially-distributed Systems}

We first generate a $30$th order discrete time system using the procedure in \cite{COL12} with $5$ poles with the largest modulus lying in $[0.8,0.9]$ and one pole with the $6$th largest modulus smaller than $0.75$. For convenience, we let ${\tt p}_{k}$, whose real part is ${\tt a}_{k}$ and the imaginary part ${\tt b}_{k}$, denote the pole of this system with the $k$th largest modulus. Then we generate $M=500$ new test systems by keeping zeros and poles of this system unchanged except the $5$ poles with the largest modulus, i.e., $\{{\tt p}_{k}\}_{k=1}^{5}$ and then. Note that if there exists an unpaired non-real pole in $\{{\tt p}_{k}\}_{k=1}^{5}$, e.g., ${\tt a}_{k}+{\tt b}_{k}\mathrm{i}$ with ${\tt b}_{k}\neq 0$ is included but ${\tt a}_{k}-{\tt b}_{k}\mathrm{i}$ is not, we will regenerate the original system until $\{{\tt p}_{k}\}_{k=1}^{5}$ include either real poles or complex conjugate pairs of poles. For the $i$th new test system with $i=1,\cdots,500$, $\{{\tt p}_{k}\}_{k=1}^{5}$ are modified as $\{{\tt p}_{k,i}\}_{k=1}^{5}$ with the real part ${\tt a}_{k,i}$ and the imaginary part ${\tt b}_{k,i}$ as follows,
\begin{itemize}
	\item[-] for real pole ${\tt p}_{k}$, the modified ${\tt p}_{k,i}={\tt a}_{k,i}$ is uniformly distributed in $[{\tt a}_{k}-0.05,{\tt a}_{k}+0.05]$ and ${\tt b}_{k,i}$=0;
	\item[-] for complex conjugate pair of poles ${\tt p}_{k}={\tt a}_{k}\pm {\tt b}_{k}\mathrm{i}$, the modified complex conjugate pair of poles are ${\tt a}_{k,i}\pm{\tt b}_{k,i}\mathrm{i}$, where $({\tt a}_{k,i},{\tt b}_{k,i})$ is uniformly distributed in the circle with the center $({\tt a}_{k},{\tt b}_{k})$ and radius $0.05$. 
\end{itemize} 
Now we obtain $500$ spatially-distributed test subsystems and for the $i$th subsystem, the corresponding location is $p_{i}=[{\tt a}_{1,i}, {\tt b}_{1,i}, \cdots,{\tt a}_{5,i},{\tt b}_{5,i}]^{T}\in\R^{10}$.

\subsubsection{Test Data Sets}

We choose the test input signal $u(t_{j})=e^{-\alpha t_{j}}\sin(\omega_{0}t_{j})$ with $\alpha=10^{-2}$ and $\omega_{0}=\pi/8$, whose state-space model is in the form of
\begin{align}
	\label{eq:ss for ut}
	\begin{array}{l}
		\tilde{z}_{i,j+1}=\tilde{E}\tilde{z}_{i,j}+\tilde{F}\delta_{j},\ \tilde{z}_{i,0}=0\in\R^{2},\\
		u(t_{j})=\tilde{H}\tilde{z}_{i,j},\ j=0,1,\cdots,N,
	\end{array}
\end{align}
where $\tilde{z}_{i,j}\in\R^{2}$, $\delta_{j}$ denotes the impulsive input, i.e., $\delta_{j}=1$ for $j=0$ and $\delta_{j}=0$ for $j=1,\cdots$, and
\begin{subequations}
	\label{eq:def of tilde_EFH}
\begin{align}
	&\tilde{E}=\left[\begin{array}{cc} 2e^{-\alpha}\cos(\omega_{0}) & -e^{-2\alpha}\\ 1 & 0\end{array}\right],\\
	&\tilde{F}=\left[\begin{array}{c} 1  \\ 0 \end{array}\right], \tilde{H}=\left[\begin{array}{c} e^{-\alpha}\sin(\omega_{0}) \\ 0 \end{array}\right]^{T}.
\end{align}
\end{subequations}
Then for $i=1,\cdots,M$, we simulate the $i$th test subsystem with the test input signal to get the noise-free output $f(p_{i},t_{j})$ and then corrupt it with an additive measurement noise $v_{i,j}$, which follows a Gaussian distribution with zero mean and variance $\sigma^2$, leading to a data record with $400$ pairs of input and measurement output data $\{y_{i,j},u(t_{j})\}_{j=1}^{400}$. The average signal-to-noise ratio (SNR) of $500$ test subsystems is $1$, where the SNR of each test subsystem is defined as the ratio between the variance of the noise-free output $f(p_{i},t_{j})$ and that of the measurement noise $v_{i,j}$. In this way,  the generated data sets contain $500$ data records, each with $400$  pairs of input and measurement output data, i.e., $\{y_{i,j},u(t_{j})\}_{i=1,j=1}^{500,400}$.

\subsubsection{Choice of Kernels}

For the ``spatial-temporal'' approach, the spatial kernel $k_{s}(p_{i},p_{i'};\alpha_{s})$ and the temporal kernel $k_{t}(t_{j},t_{j'};\alpha_{t})$ in \eqref{eq:stcovfunc1} are chosen to be the SE kernel \eqref{eq:spatialkernel}, and the following one, respectively, 
\begin{align}\label{eq:kt with kappa}
	k_{t}(t_{j},t_{j'};\alpha_{t})=\kappa(t_{j},t_{j'};\alpha_{t})\sum_{k=1}^{n_{b}}\sum_{k'=1}^{n_{b}}u(t_{j-k})u(t_{j'-k'}),
\end{align}
where $\kappa(t_{j},t_{j'};\alpha_{t}): \R_+ \times \R_+ \rightarrow\R$ is the diagonal correlated (DC) kernel in \cite{COL12}, i.e.,
	\begin{align}
	\label{eq:DC kernel}
		&\kappa(t_{j},t_{j'};\alpha_{t})=\delta_{t}\lambda_{t}^{(t_{j}+t_{j'})/2}\rho_{t}^{|t_{j}-t_{j'}|},\\
		&\alpha_{t}=[\delta_{t}, \lambda_{t},\rho_{t}]\in\Omega=\{\delta_{t}\geq 0,\lambda_{t}\in[0,1), |\rho_{t}|\leq 1 \}.\nonumber
	\end{align}

Noting the state-space model realization of the DC kernel \eqref{eq:DC kernel} in \cite{Chen18} and \eqref{eq:ss for ut}, it can be shown that the state-space model realization of the spatial-temporal kernel \eqref{eq:stcovfunc1} with \eqref{eq:spatialkernel} as the spatial kernel and \eqref{eq:kt with kappa} as the temporal kernel takes the form of \eqref{eq:NewDTstate} by replacing \eqref{eq:ss1 for NewDTstate} with
\begin{align}
	&	s_{j+1}=Fs_{j}+G_{j}w_{j},\ s_{1}\sim\mathcal{N}(0,I_{M}\otimes \Sigma_{1}),\nonumber
\end{align}
 and using
\begin{align}
	&r=3,\ F=\left[\begin{array}{cc} I_{M}\otimes (\lambda_{t}^{1/2}\rho_{t}) & 0\in\R^{M\times 2M}\\ I_{M}\otimes ((1-\rho_{t}^2)^{1/2}\tilde{F}) & I_{M}\otimes \tilde{E} \end{array}\right],\nonumber\\
	&G_{j}=\left[\begin{array}{c} (I_{M}\otimes \lambda_{t}^{1/2})\delta_{t}^{1/2}\lambda_{t}^{t_{j}/2}\\ 0\in\R^{2M\times M} \end{array}\right],\ H=\left[\begin{array}{c}0\in\R^{M\times M} \\ (I_{M}\otimes \tilde{H}) \end{array} \right]^{T},\nonumber\\
	&\Sigma_{1}=\left[\begin{array}{cc}I_{M}\otimes (\delta_{t}/(1-\rho_{t}^2)) & 0\in\R^{M\times 2M}\\ 0\in\R^{2M\times M} & 0\in\R^{2M\times 2M}  \end{array}\right],\nonumber
\end{align}
where $\tilde{E},\tilde{F}$ and $\tilde{H}$ are given in \eqref{eq:def of tilde_EFH}. Note that for the ``temporal'' approach, we only apply the DC kernel \eqref{eq:DC kernel}.

\subsubsection{Hyper-parameter estimation and Impulse Response Estimation}

For the ``spatial-temporal'' approach, we use the MLM method \eqref{eq:loglikelihood} to estimate $\alpha=[\alpha_{s}^{T},\alpha_{t}^{T},\sigma^2]^{T}$ and apply the same strategy as stated in Section \ref{subsubsec:hyper-parameter est} for finding a ``good'' local minimum. With the estimated hyper-parameter, we further run the Kalman filter and smoother to obtain the estimates of $\{b_{i}(p_{i},t_{j})\}_{i=1,j=1}^{M,n_{b}}$, denoted as $\{\hat{b}_{i,j|N}\}_{i=1,j=1}^{M,n_{b}}$, where for $j=1,\cdots,n_{b}$, 
\begin{align}
\hat{b}_{j|N}=\left[\begin{array}{ccc}\hat{b}_{1,j|N} & \cdots & \hat{b}_{M,j|N} \end{array} \right]^{T}\in\R^{M},
\end{align}
can be obtained by
\begin{align}
	\hat{b}_{j|N}=\Lambda D^{1/2}(I_{M}\otimes (1-\rho_{t}^2)^{1/2})[\hat{x}_{j|N}]_{1:M}.
\end{align}
Here $\Lambda$ and $D$ are defined in \eqref{eq:SVDofKs}, and $[\hat{x}_{j|N}]_{1:M}$ denotes a vector containing the first $M$ elements of $\hat{x}_{j|N}$ in \eqref{eq:hatx1}.

For the ``temporal'' approach, we use the MLM method \eqref{eq:loglikelihood} to estimate $\alpha=[\alpha_{t}^{T},\sigma^2]^{T}$ and then with the estimated hyper-parameter, we calculate $\hat{b}_{i}=[\hat{b}_{i,1|N},\cdots,\hat{b}_{i,n_{b}|N}]^{T}\in\R^{n_{b}}$ for the $i$th system with $i=1,\cdots,M$, where the implementation \cite{CL13} is used.

To evaluate the estimation performance of $\{\hat{b}_{i,j|N}\}_{i=1,j=1}^{M,n_{b}}$, for the $i$th system with $i=1,\cdots,M$, we let
\begin{align*}
	b_{i}^{0}=&\left[\begin{array}{ccc}b_{i,1}^{0} & \cdots & b_{i,n_{b}}^{0}\end{array} \right]^{T},\
\end{align*}
denote the true value of $[b_{i}(p_{i},t_{1}),\cdots,b_{i}(p_{i},t_{n_{b}})]^{T}$, and then define the measure of fit, e.g., \cite{Ljung:00},
\begin{align*}
	\text{fit}_{i}^{b}=100\times\left(1- \frac{\|\hat{b}_{i}-b_{i}^{0}\|_{2}}{\|b_{i}^{0}-\bar{b}_{i}^{0}\|_{2}}\right),\ \bar{b}_{i}^{0}=\frac{1}{n_{b}}\sum_{j=1}^{n_{b}}b_{i,j}^{0}.
\end{align*}
The average estimation fit of $\{\hat{b}_{i}\}_{i=1}^{M}$ is defined as
\begin{align}\label{eq:def of bar_fitg}
	\overline{\text{fit}}^{b}=\frac{1}{M}\sum_{i=1}^{M}\text{fit}_{i}^{b}.
\end{align}

\subsubsection{Simulation Results and Findings}

In the simulation, we choose the FIR order $n_b=125$. The average estimation fits of $\{\hat{b}_{i}\}_{i=1}^{M}$ of the ``spatial-temporal'' and ``temporal'' approaches in Table \ref{table: average fits} show that the ``spatial-temporal'' approach gives much better estimation performance than the ``temporal'' approach. This observation indicates that exploring the spatial interconnections  among subsystems is beneficial for the identification of spatially-distributed system.



\begin{table}[!htb]
	\begin{center}
		
		\caption{Average estimation fits of $M$ spatially-distributed systems}
		\label{table: average fits}\vspace*{2mm}
		\begin{tabular}{ccc}
			\hline
			Approach & ``spatial-temporal''  & ``temporal'' \\
			\hline
			$\overline{\text{fit}}^{b}$ \eqref{eq:def of bar_fitg} & 77.27 & 8.78  \\ 
			\hline 
		\end{tabular}
	\end{center}
\end{table}

	\section{Conclusion}\label{sec:6Conclusion}
	In this paper, we proposed an efficient implementation with computational complexity $\mathcal{O}(M^3+NM^2)$, for spatial-temporal Gaussian process regression by exploring the Kronecker structure of its state-space model realization, where $N$ and $M$ are the numbers of time instants and locations, respectively. The proposed implementation has been illustrated over applications in weather data prediction and spatially-distributed system identification. For the weather prediction, the design kernel is shown to give better prediction performance than the one in \cite{TCCPS20} and for the spatially-distributed system identification, the benefit of exploring the spatial interconnections among subsystems is confirmed. 



	\def\thesectiondis{\thesection.}                   
	\def\thesubsectiondis{\thesection.\arabic{subsection}.}          
	\def\thesubsubsectiondis{\thesubsection.\arabic{subsubsection}.}
	
	\setcounter{subsection}{0}
	
	\renewcommand{\thesection}{A}
	\setcounter{theorem}{0}

	\renewcommand{\theequation}{A.\arabic{equation}}
	\setcounter{equation}{0}
	
	\renewcommand{\thesubsection}{\thesection.\arabic{subsection}}

	\section*{Appendix A}\label{sec:Appendix A}
	This appendix contains the proofs of all theoretical results and the derivations of state-space model of \eqref{eq:Exp2tp}.
	
	%
	\subsection{Proof of Proposition \ref{prop:diagonal structure of Psi}}
	

	According to \eqref{eq:linear representation form of innovation for j=1}-\eqref{eq:linear representation form of innovation for j>1} in Lemma \ref{lemma:preliminary results of the innovation sequence} and \eqref{eq:def of Theta}, we have
	\begin{align} \label{eq:key}
		\Theta = 
		\begin{bmatrix}
			l_1 \\ l_{2}-b_{2,1}l_{1}\\ \vdots \\ l_N - \sum_{i=1}^{N-1}b_{N,i}l_{i}
		\end{bmatrix} = \Gamma L,
	\end{align}
	where $\Gamma$ is defined in \eqref{eq:def of Gamma}.
	Then, inserting \eqref{eq:equality of Theta and Gamma} into \eqref{eq:def of Psi}, it follows that $\Psi=\mathbb{COV}[\Gamma L, \Gamma L]
		=\Gamma\mathbb{COV}[L,L]\Gamma^{T}$, which leads to \eqref{eq:Psi using Gamma} using \eqref{eq:def of Sigma_bar}. Combining \eqref{eq:def of E_bar_j}, \eqref{eq:def of Psi} and \eqref{eq:uncorrelatedness of innovation}, we can obtain \eqref{eq:Psi block diagonal matrix}.
	
	\subsection{Proof of Proposition \ref{prop:computation of MLM cost function}}
	
First, note that the computation of the cost function of  \eqref{eq:loglikelihood} depends on that of $\log|\Sigma(\alpha)| $ and $Y^T\Sigma^{-1}(\alpha)Y$. Then  following the idea of \cite{BP18}, where the computation of the generalized cross validation filter is discussed, and using \eqref{eq:relation2} and Proposition \ref{prop:diagonal structure of Psi},  $\log|\Sigma(\alpha)| $ and $Y^T\Sigma^{-1}(\alpha)Y$ can be computed as follows 
	\begin{subequations} 
		\begin{align}
			\label{eq:logSigma derivation}
			&\log |\Sigma(\alpha)| 
			\nonumber\\ 
			=&  \log|(I_N \otimes\Lambda) \overline{\Sigma}(\alpha) (I_N \otimes\Lambda^T)|
\\ 
			=&  \log|(I_N \otimes\Lambda)(I_N \otimes\Lambda^T)|+ \log|\overline{\Sigma}(\alpha)|
			=  \log|\Psi| \nonumber\\
			&Y^T\Sigma^{-1}(\alpha)Y
			\nonumber\\ =& L^T(I_N \otimes\Lambda^T)(I_N \otimes\Lambda)\overline{\Sigma}(\alpha)^{-1} (I_N \otimes\Lambda^T)(I_N \otimes\Lambda)L
			\nonumber\\ =& (\Gamma^{-1} \Theta)^T (\Gamma^T \Psi \Gamma)^{-1} (\Gamma^{-1} \Theta) 
			 = \Theta^T \Psi^{-1} \Theta, \label{eq:YinvSigmaY derivation}
		\end{align}
	\end{subequations}
    where the first steps of both \eqref{eq:logSigma derivation} and \eqref{eq:YinvSigmaY derivation} are derived from \eqref{eq:relation2}, and the second step of \eqref{eq:YinvSigmaY derivation} is derived from \eqref{eq:equality of Theta and Gamma} and \eqref{eq:Psi using Gamma}. Then using \eqref{eq:def of Theta} and \eqref{eq:Psi block diagonal matrix}, we can obtain \eqref{eq:relation3}.

   \subsection{Proof of Proposition \ref{prop:computation of GCV and SURE cost function}}
   
   As shown in \eqref{eq:GCV} and \eqref{eq:SURE}, the computation of the cost functions of the GCV and SURE methods depends on that of $\delta$ and $S$ defined in \eqref{eq:def of delta} and \eqref{eq:def of S}, respectively. For convenience, we let $\gamma = \sigma^2$. We first rewrite \eqref{eq:Sigma} as 
	\begin{align} \label{eq:ReSigma}
		\gamma\Sigma(\alpha)^{-1} = I_{NM} - \left[K_t(\alpha_t)\otimes K_s(\alpha_s)\right]\Sigma(\alpha)^{-1}.
	\end{align}
	Following the discussions in \cite{BP18}, we can represent $\delta$ and $S$ as functions of $\log|\Sigma(\alpha)| $ and $Y^T\Sigma^{-1}(\alpha)Y$, respectively,
	\begin{align}
		&\gamma\frac{\partial \log |\Sigma(\alpha)|}{\partial \gamma} \nonumber\\ &= \gamma \trace(\Sigma(\alpha)^{-1} \frac{\partial \Sigma(\alpha)}{\partial \gamma}) = \gamma \trace(\Sigma(\alpha)^{-1}) \nonumber\\
		& = \trace\left\{ I_{NM} - \left[ K_t(\alpha_t)\otimes K_s(\alpha_s)\right]\Sigma(\alpha)^{-1}\right\} \nonumber\\
		& =  NM-\trace \left\{  \left[ K_t(\alpha_t)\otimes K_s(\alpha_s)\right]\Sigma(\alpha)^{-1} \right\} \nonumber\\
		& =  NM- \delta  ,
		\\
		&-\gamma^2 \frac{\partial Y^T\Sigma(\alpha)^{-1}Y}{\partial \gamma} \nonumber\\
		&=\gamma^2 Y^T \Sigma^{-1}\frac{\partial \Sigma(\alpha)}{\partial \gamma}\Sigma^{-1} Y 
		=\gamma^2 Y^T \Sigma^{-2} Y \nonumber\\
		&=\gamma^2 Y^T \left\{ I_{NM} - \left[ K_t(\alpha_t)\otimes K_s(\alpha_s)\right] \Sigma(\alpha)^{-1} \right\}^T \nonumber\\ &\quad \left\{ I_{NM} - \left[ K_t(\alpha_t)\otimes K_s(\alpha_s)\right]\Sigma(\alpha)^{-1}\right\}  Y\nonumber\\
		& =||\hat{Y}-Y||^2_2=S.
	\end{align}
	Then by using \eqref{eq:relation3},  $S$ and $\delta$ can be computed as follows
		\begin{align}\label{eq:recursivenewSdelta}
			S &=-\gamma^2 \frac{\partial Y^T\Sigma(\alpha)^{-1}Y}{\partial \gamma}=-\gamma^2 \sum_{j=1}^N \frac{\partial \bar{e}_j^T\bar{E}_j^{-1}\bar{e}_j}{\partial \gamma},\nonumber\\
			\delta &= MN -\gamma\frac{\partial \log |\Sigma(\alpha)|}{\partial \gamma}= MN - \gamma \sum_{j=1}^N\frac{\partial\log |\bar{E}_j|}{\partial \gamma}.
		\end{align}
	Now we define
	\begin{align} \label{eq:Zeta}
		&\bar{\zeta}_{j|j-1} =\frac{\partial \hat{x}_{j|j-1}}{\partial \gamma} , 
		\bar{P}_{j|j-1} =\frac{\partial \overline{\Sigma}_{j|j-1}}{\partial \gamma},
	\end{align}
	and then ${\partial \bar{e}_j^T\bar{E}_j^{-1}\bar{e}_j}/{\partial \gamma}$ 
	and ${\partial\log |\bar{E}_j|}/{\partial \gamma}$ in \eqref{eq:recursivenewSdelta} can be further expressed as 
	\begin{subequations}
		\label{eq:eEe and logEk}
	\begin{align}
		\label{eq:eEe}
		&-\frac{\partial \bar{e}_j^T\bar{E}_j^{-1}\bar{e}_j}{\partial \gamma} 
		=\bar{e}_j^T \bar{E}_j^{-1} (\bar{H} \bar{P}_{j|j-1} \bar{H}^T + I_M)
		\bar{E}_j^{-1} \bar{e}_j ,
		\nonumber\\&\qquad\qquad\qquad\quad+ 2\bar{\zeta}_{j|j-1}^T \bar{H}\bar{E}_j^{-1} \bar{e}_j ,\\ 
		\label{eq:logEk}
		&\frac{\partial \log |\bar{E}_j|}{\partial \gamma} 
		={\rm trace} \left[ \bar{E}_j^{-1}(\bar{H} \bar{P}_{j|j-1} \bar{H}^T + I_M)\right].
	\end{align}
\end{subequations}
   Combining \eqref{eq:recursivenewSdelta} with \eqref{eq:eEe and logEk}, we can obtain \eqref{eq:S and delta}. Moreover, inserting \eqref{eq:Nxkk} and \eqref{eq:Nxk} into \eqref{eq:Zeta}, and  \eqref{eq:NSigmakk} and \eqref{eq:NSigma} into \eqref{eq:Zeta}, we can compute $\bar{\zeta}_{j|j-1}$ and $\bar{P}_{j|j-1}$ recursively as shown in Proposition \ref{prop:computation of GCV and SURE cost function}.

	\subsection{Proof of Theorem \ref{thm:computational complexity}}\label{subsec:AppendixThm1}
	
 As shown in Algorithm \ref{alg:The Proposed Implementation}, the proposed implementation consists of three steps, and
	 in what follows, we will study their computational complexities, respectively:
	\begin{enumerate}[1)]
		\item {\it Computational complexity of Step 1:} 
		We first calculate the SVD of $K_{s}\in\R^{M\times M}$ in \eqref{eq:SVDofKs} and its computational complexity is $\mathcal{O}(M^3)$. Then the computational complexities of \eqref{eq:F_bar H_bar} and \eqref{eq:Loutput} are $\mathcal{O}(Mr)$ and $\mathcal{O}(NM^2)$, respectively. Hence, this step has the computational complexity $\mathcal{O}(M^3+NM^2)$.
		
		\item {\it Computational complexity of Step 2:}
		Since the evaluation of the cost functions of three hyper-parameter estimation methods all rely on Kalman filter, we first consider the computational complexity of the Kalman filter \eqref{eq:NewKalmanFilter} and then that of three hyper-parameter estimation methods, respectively.
		
		\begin{enumerate}
			\item {\it Computational complexity of Kalman filter \eqref{eq:NewKalmanFilter}}: To show the computational complexity of the Kalman filter, we first use induction to show that, for $j=1,\cdots,N$, $\bar E_j\in\R^{M\times M}$ and $\overline{\Sigma}_{j|j-1}\in\mathbb{R}^{Mr \times Mr}$ are diagonal and block diagonal matrices, respectively. It consists of two steps.
			
			Our first step is to prove that $\bar E_1$ and $\overline{\Sigma}_{1|0}$ are diagonal and block diagonal matrices, respectively. For $\overline\Sigma_{1|0}$, inserting \eqref{eq:F_bar H_bar}, \eqref{eq:def of Q} and
			$
				\overline{\Sigma}_{0|0}
				=\E[ (x_{0}-\E(x_{0}))(x_{0}-\E(x_{0}))^{T}]=I_{M}\otimes \Sigma_{0},
			$
		    where we apply \eqref{eq:stateupdate} and \eqref{eq:def of hat_xjm and overline_Sigmajm}, into \eqref{eq:NSigma}, we have
		    \begin{align}\label{eq:overline_Sigma10}
		    	\overline{\Sigma}_{1|0} 
		    	&= I_M \otimes (F_D\Sigma_0F_D^T + G_DG_D^T)\nonumber\\
		    	&=\text{blkdiag}(\bar{\Sigma}_{1,1},\cdots,\bar{\Sigma}_{1,M}),
		    \end{align}
	        where $	\bar{\Sigma}_{1,i}=F_D\Sigma_0F_D^T + G_DG_D^T$ for $i=1,\cdots,M$. For $\bar{E}_1$, we insert \eqref{eq:F_bar H_bar} and \eqref{eq:overline_Sigma10} into \eqref{eq:def of E_bar_j} to obtain
		   \begin{align}\label{eq:barE1}
		   	\bar{E}_1 
		   	&=H_D \bar{\Sigma}_{1,1} H_D^T D + \sigma^2I_M\nonumber\\
		   	&=\text{diag}(\bar{E}_{1,1},\cdots,\bar{E}_{1,M}),
		   \end{align}
	       where $\bar{E}_{1,i}=[D]_{ii}H_D \bar{\Sigma}_{1,1} H_D^T+\sigma^2$ for $i=1,\cdots,M$.
       
           Our second step is to show that for $j=1,\cdots,N-1$, if we assume $\bar E_j$ and $\overline{\Sigma}_{j|j-1}$ are diagonal and block diagonal matrices, respectively, then we can show that $\bar E_{j+1}$ and $\overline{\Sigma}_{j+1|j}$ are diagonal and block diagonal matrices, respectively. 
           For convenience, we define that
           \begin{subequations}\label{eq:def of bar_Ej and overline_Sigmajj-1}
           \begin{align}
           	\bar{E}_{j}=&\text{diag}(\bar{E}_{j,1},\cdots,\bar{E}_{j,M}),\\
           	\label{eq:def of overline_Sigmajj-1}
           	\overline{\Sigma}_{j|j-1}=&\text{blkdiag}(\bar{\Sigma}_{j,1},\cdots,\bar{\Sigma}_{j,M}),
           \end{align}
          \end{subequations}
           where $\bar{E}_{j,i}\in\R$ and $\bar{\Sigma}_{j,i}\in\R^{r\times r}$ for $i=1,\cdots,M$.
           Combining \eqref{eq:NSigma} and \eqref{eq:NSigmakk}, we have
           \begin{align}\label{eq:def of overline_Sigmaj+1j}
           \overline{\Sigma}_{j+1|j}=&\bar{F}\overline{\Sigma}_{j|j-1}\bar{F}^{T}+Q\nonumber\\
           	& -\bar{F}\overline{\Sigma}_{j|j-1} \bar{H}^T \bar{E}_j^{-1}\bar{H}\overline{\Sigma}_{j|j-1} \bar{F}^{T}\nonumber\\
           	=&\text{blkdiag}(\bar{\Sigma}_{j+1,1},\cdots,\bar{\Sigma}_{j+1,M}),\\
           	\bar{\Sigma}_{j+1,i}=&F_{D}\bar{\Sigma}_{j,i}F_{D}^{T}+G_{D}G_{D}^{T}\\
           	&+([D]_{ii}/\bar{E}_{j,i})F_{D}\bar{\Sigma}_{j,i}H_{D}^{T}H_{D}\bar{\Sigma}_{j,i}F_{D}^{T},\nonumber
           	\end{align}
           where $i=1,\cdots,M$, and we apply \eqref{eq:F_bar H_bar}, \eqref{eq:def of Q}, \eqref{eq:def of bar_Ej and overline_Sigmajj-1} and the fact that $\bar{H}^T \bar{E}_j^{-1}\bar{H}=(I_{M}\otimes H_{D}^{T})[(D^{1/2}\bar{E}_{j}^{-1}D^{1/2})\otimes 1](I_{M}\otimes H_{D})=(D^{1/2}\bar{E}_{j}^{-1}D^{1/2})\otimes(H_{D}^{T}H_{D})$ is a block diagonal matrix with $i$th block being $([D]_{ii}/\bar{E}_{j,i})H_{D}^{T}H_{D}\in\R^{r\times r}$.
           Then for $\bar{E}_{j+1}$, we use \eqref{eq:F_bar H_bar} and \eqref{eq:def of overline_Sigmaj+1j} to obtain
           \begin{align}
           	\bar{E}_{j+1}=&\bar{H}\overline{\Sigma}_{j+1|j}\bar{H}^{T}+\sigma^2 I_{M}\nonumber\\
           	=&\text{diag}(\bar{E}_{j+1,1},\cdots,\bar{E}_{j+1,M}),
           \end{align}
           where $\bar{E}_{j+1,i}=[D]_{ii}H_{D}\bar{\Sigma}_{j+1,i}H_{D}^{T}+\sigma^2$ for $i=1,\cdots,M$.
           
           Hence for $j=1,\cdots,N$, it is clear that $\bar E_j\in\R^{M\times M}$ and $\overline{\Sigma}_{j|j-1}\in\mathbb{R}^{Mr \times Mr}$ are diagonal and block diagonal matrices, respectively.  It follows that $\overline{\Sigma}_{j|j}\in\R^{Mr\times Mr}$ in \eqref{eq:NSigmakk} is also a block diagonal matrix due to that $\bar{H}^T \bar{E}_j^{-1}\bar{H}$ is a block diagonal matrix.

			Then, according to the properties of the Kronecker product and the matrix multiplication, for each $j=1,\cdots,N$, the computational complexities of \eqref{eq:def of E_bar_j}, \eqref{eq:NEk} and \eqref{eq:Nxkk}, and \eqref{eq:NSigmakk} and \eqref{eq:NSigma} are $\mathcal{O}(Mr^2)$ and $\mathcal{O}(Mr^3)$, respectively. Thus the computational complexity of \eqref{eq:NewKalmanFilter} is $\mathcal{O}(NM)$. 
			
			\item {\it Computational complexity of the computation of the cost function of the MLM method:}
			As shown in Proposition \ref{prop:computation of MLM cost function}, to calculate the cost function of the MLM method \eqref{eq:loglikelihood}, we first calculate $\bar{e}_j$ and $\bar{E}_{j}$ as shown in \eqref{eq:NewKalmanFilter} for $j=1,\cdots,N$ and the computational complexity is $\mathcal{O}(NMr^3)$. Then we calculate \eqref{eq:relation3} and \eqref{eq:loglikelihood}, whose computational complexity is $\mathcal{O}(NM)$.	Therefore, 
			the computational complexity of the cost function of the MLM method is $\mathcal{O}(NM)$.
			
			\item {\it Computational complexities of the computation of the cost functions of the GCV and SURE methods:}
			For the GCV method \eqref{eq:GCV} and the SURE method \eqref{eq:SURE}, as shown in Proposition \ref{prop:computation of GCV and SURE cost function}, since $\bar{E}_{j}$ is a diagonal matrix, and $\overline{\Sigma}_{j|j-1}$, $\bar{F}$ and $\bar{H}$ are block diagonal matrices, the computational complexity of $\bar{\zeta}_{j|j-1}\in\mathbb{R}^{Mr }$ and $\bar{P}_{j|j-1}\in\mathbb{R}^{Mr \times Mr}$ with $j=1,\cdots,N$ in \eqref{eq:Nzeta} and \eqref{eq:NPk} are $\mathcal{O}(Mr^3)$. Therefore, 
			the computational complexities of the cost functions of the GCV and the SURE methods are both $\mathcal{O}(NM)$.
			
		\end{enumerate}
		
		\item {\it Computational complexity of Step 3:} We discuss the computational complexities of the Kalman smoother \eqref{eq:NewKalmanSmoother} and predictor  \eqref{eq:NewKalmanPredictor}, respectively.
		
		\begin{enumerate}
		
		\item {\it Computational complexity of Kalman smoother:}
		Since $\overline{\Sigma}_{j|j}\in\R^{Mr\times Mr}$, $j=N-1,\cdots,1$ are block diagonal matrices, $\bar{J}_j$, $j=N-1,\cdots,1$ are also block diagonal matrices. Hence, for each $j$, the computational complexities of \eqref{eq:NSigmajj}, and \eqref{eq:barJ1}, \eqref{eq:hatx1} and \eqref{eq:NSigmakN1} are $\mathcal{O}(Mr^2)$ and $\mathcal{O}(Mr^3)$, respectively. Finally, since the output transform in \eqref{eq:NfkN1} has computational complexity $\mathcal{O}(M^2r)$,  the computational complexity of \eqref{eq:NewKalmanSmoother} is $\mathcal{O}(NM^2)$.
		
		\item {\it Computational complexity of Kalman predictor:}
		For each $j$, the computational complexities of \eqref{eq:NSigmakN3} and \eqref{eq:hatf2} are $\mathcal{O}(Mr^3)$ and $\mathcal{O}(M^2r)$, respectively. Thus the computational complexity of \eqref{eq:NewKalmanPredictor} is $\mathcal{O}(NM^2)$.
		
		\end{enumerate}
	\end{enumerate}
	Hence, the proof of Theorem  \ref{thm:computational complexity} is complete.
	\subsection{State-space Model Realization of \eqref{eq:Exp2tp}}\label{subsec:state-space realization of kernel^tpe}  
	The kernel \eqref{eq:Exp2tp} can be divided into three parts:
\tiny		\begin{align}\label{eq:Exp2tp3kernels}
		&k_{\text{Te2}}(\tau;\delta_{t},c_{t})k_{\text{EXP}}(\tau)\\ 
		&=\delta_t [ \underbrace{(1-c_t+\frac{3}{4}c_t^2)\exp\left(-\frac{|\tau|}{\sigma_t}\right)}_{\text{(\ref{eq:Exp2tp3kernels}a)}}\nonumber
		\\ \nonumber 
		&+\underbrace{(c_t-c_t^2)\cos(2\pi \texttt{f}|\tau|)\exp\left(-\frac{|\tau|}{\sigma_t}\right)}_{\text{(\ref{eq:Exp2tp3kernels}b)}}   \\ \nonumber 
		&+\underbrace{\frac{c_t^2}{4}\cos(4\pi \texttt{f}|\tau|)\exp\left(-\frac{|\tau|}{\sigma_t}\right)}_{\text{(\ref{eq:Exp2tp3kernels}c)}} ],
	\end{align}\normalsize 
	where (\ref{eq:Exp2tp3kernels}a) is an exponential kernel, and (\ref{eq:Exp2tp3kernels}b) and (\ref{eq:Exp2tp3kernels}c) are periodic kernels with different periods. 
	To obtain the state-space model of \eqref{eq:Exp2tp}, we derive below the state-space models of these kernels, respectively.
	
	Firstly, we denote the PSD of (\ref{eq:Exp2tp3kernels}a) as $\Phi_{\text{EXP}}(\omega)$, which can be obtained using \eqref{eq:PSD} as follows
	\begin{align}
		\label{eq:PSDofEXP}
		&\Phi_{\text{EXP}}(\omega) =  \frac{\delta_t (1-c_t+\frac{3}{4}c_t^2)(1-e^{-2\beta_{t}})}{\left(e^{- {\im \omega}}-e^ {-\beta_{t}}\right)	\left(e^{ {\im \omega}}-e^{- \beta_{t}}\right)},  
	\end{align}
	where $\beta_{t}=\frac{1}{\sigma_t}$.	According to Assumption \ref{ass:timespacekernl} and \eqref{eq:factorization}, we can obtain the transfer function of \eqref{eq:PSDofEXP} in the form 
	\begin{align} \label{eq:transferfunctionEXP}
		W_{\text{EXP}}(e^{\im\omega}) = \frac{\sqrt{ \delta_t(1-c_t+\frac{3}{4}c_t^2) (1-e^{ -2\beta_{t}})} } {e^{{\im \omega}}-e^{ -\beta_{t}}}.
	\end{align}
	
Then we can derive the corresponding state-space model using the realization theory in \cite{Chen99}
		\begin{subequations}\label{eq:state-space models 1} as follows
			\begin{align}
				s_{1,j+1}&=F_{1}s_{1,j}+G_{1}w_{1,j}, j=1,\cdots \\ y_{1,j}&=H_{1,j}s_{1,j}, 
			\end{align}
		\end{subequations}
		where $s_{1,{j}}\in\R$, $F_1 =e^{ -\beta_{t}}$, $G_1 = 1$ and $H_{1,j} = \sqrt{\delta_t(1-c_t+\frac{3}{4}c_t^2) (1-e^{ -2\beta_{t}}) }$ and $w_{1,{j}}\in\R$ is white Gaussian noise with zero mean and unit variance.
	
	Secondly, we denote the PSD of (\ref{eq:Exp2tp3kernels}b) as $\Phi_{\text{PD1}}(\omega)$ and it can be derived using \eqref{eq:PSD} as follows
	\begin{align}
		\label{eq:PSDofPD1}
		&\Phi_{\text{PD1}}(\omega)= \delta_t (c_t-c_t^2) \\
		&\left[ \frac{1}{2} \left(e^{3\beta_{t}}-e^{ \beta_{t}}\right)\left(e^{- {\im\varrho_{1} }}+e^{ {\im \varrho_{1} }}\right) 
		\left(e^{- {\im \omega}}+e^{ {\im \omega}}\right)  +1-e^{4 \beta_{t}}\right]   \nonumber \\
		& / \left[ \left(e^{\beta_{t}}-e^{- {\im \varrho_{1} }- {\im \omega}}\right)\left(e^{\beta_{t}}-e^{ {\im \varrho_{1} }- {\im \omega}}\right)  \nonumber \right.\\\nonumber &\left. \left(e^{\beta_{t}}-e^{ {\im \omega}- {\im\varrho_{1}}}\right)\left(e^{\beta_{t}}-e^{ {\im\varrho_{1} }+ {\im \omega}}\right) \right], \nonumber 
	\end{align}
	where $\varrho_{1} = 2\pi \texttt{f} $.
	According to Assumption \ref{ass:timespacekernl} and \eqref{eq:factorization}, we consider the transfer function of \eqref{eq:PSDofPD1} in the form 
	\begin{align} \label{eq:transferfunction of PD1}
		W(e^{\im\omega}) =\sqrt{\delta_t (c_t-c_t^2)} \frac{n_1e^{\im\omega}+n_2}
		{e^{2{\im \omega}}- d_{2}  e^{{\im \omega}} -d_{1}},
	\end{align}
	with $d_{1} = -e^{-2 \beta_{t}}$, $d_{2} = e^{-\beta_{t}}(e^{\im \varrho_{1}}+e^{-\im \varrho_{1}})$,
	\begin{subequations}
		\begin{align}
			A &= n_1^2+n_2^2 = 1-e^{4 \beta_{t}},\\  
			B &= n_1n_2 = \frac{1}{2} \left(e^{3\beta_{t}}-e^{ \beta_{t}}\right)\left(e^{- {\im\varrho_{1} }}+e^{ {\im \varrho_{1} }}\right).
		\end{align}
	\end{subequations}
	Therefore, we can derive $n_1$ and $n_2$ by solving 
	\begin{align} \label{eq:Polynomials n1n2}
		n_1^4 - An_1^2 + B^2 = 0.
	\end{align}
		With \eqref{eq:transferfunction of PD1}, $n_1$ and $n_2$, we can derive the   state-space model in controllable canonical form of (\ref{eq:Exp2tp3kernels}b) using the realization theory in \cite{Chen99} as follows
		\begin{subequations}\label{eq:state-space models 2} 
			\begin{align}
				s_{2,j+1}&=F_{2}s_{2,j}+G_{2}w_{2,j},j=1,\cdots\\ 
				y_{2,j} &=H_{2,j}s_{2,j},
			\end{align}
		\end{subequations} where $s_{2,{j}}\in\R^2$, $w_{2,{j}}\in\R^{2}$ is white Gaussian noise with zero mean and covariance matrix $I_{2}$, and
		\begin{align}\label{eq:system matrices}
			F_{2}=\begin{bmatrix}
				0 & 1 \\
				d_{1}  & d_{2} 
			\end{bmatrix}, G_{2}=\begin{bmatrix}
				0 \\
				1 
			\end{bmatrix}, H_{2,j} =\sqrt{\delta_t (c_t-c_t^2)} \begin{bmatrix} n_{2} & n_{1}\end{bmatrix} . \nonumber
		\end{align}
		Moreover, it is necessary to check if the $n_1$ and $n_2$ guarantees that the zeros of \eqref{eq:transferfunction of PD1} are inside the unit circle and the largest eigenvalue of the corresponding system matrices $F_{2}$ should be less than 1.
	
	The PSD of (\ref{eq:Exp2tp3kernels}c) can be obtained by in a similar way as (\ref{eq:Exp2tp3kernels}b) by replacing the amplitude with $\delta_t \frac{c_t^2}{4}$ and letting $\varrho_{1}=4\pi \texttt{f}$ in \eqref{eq:PSDofPD1}, respectively. Let the state-space model of (\ref{eq:Exp2tp3kernels}c) in controllable canonical form be represented as
	\begin{subequations}\label{eq:state-space models 3}
		\begin{align}
			s_{3,j+1}&=F_{3}s_{3,j}+G_{3}w_{3,j},j=1,\cdots\\ 
			y_{3,j} &=H_{3,j}s_{3,j},
		\end{align}
	\end{subequations}
	where $s_{3,{j}}\in\R^{2}$, $F_3\in\R^{2\times 2}$, $G_3\in\R^{2}$, $H_{3,j}\in\R^{1\times 2}$, and $w_{3,{j}}\in\R^{2}$ is white Gaussian noise with zero mean and covariance matrix $I_{2}$. 
	
	Therefore, the state-space model of \eqref{eq:Exp2tp} can be obtained by combining the state-space models \eqref{eq:state-space models 1}, \eqref{eq:state-space models 2} and \eqref{eq:state-space models 3} as follows	
	\begin{subequations}\label{eq:combination}
		\begin{align*}
			&\begin{bmatrix}
				s_{1,{j+1}} \\ s_{2,{j+1}} \\ s_{3,{j+1}}
			\end{bmatrix} =\begin{bmatrix}
				F_1 & 0 & 0 \\ 0 & F_2 & 0 \\ 0 & 0 & F_3
			\end{bmatrix} \begin{bmatrix}
				s_{1,{j}} \\ s_{2,{j}} \\ s_{3,{j}}
			\end{bmatrix} + \begin{bmatrix}
				G_1 & 0 & 0 \\ 0 & G_2 & 0 \\ 0 & 0 & G_3 \end{bmatrix}
			\begin{bmatrix}
				w_{1,{j}} \\ w_{2,{j}} \\ w_{3,{j}}
			\end{bmatrix}, \\
			&y_j = \begin{bmatrix}
				H_{1,{j}} & H_{2,{j}} & H_{3,{j} }
			\end{bmatrix}\begin{bmatrix}
				s_{1,{j}}^T & s_{2,{j}}^T & s_{3,{j}}^T
			\end{bmatrix}^T.
		\end{align*}
	\end{subequations}	
	
	
	
	
	%
	%
	
	\def\thesectiondis{\thesection.}                   
	\def\thesubsectiondis{\thesection.\arabic{subsection}.}          
	\def\thesubsubsectiondis{\thesection.\thesubsection.\arabic{subsubsection}.}
	
	\renewcommand{\thesection}{B}
	\renewcommand{\thesubsection}{B.\arabic{subsection}}
	\setcounter{subsection}{0}

	\renewcommand{\theequation}{B.\arabic{equation}}
	\setcounter{equation}{0}

	\bibliographystyle{abbrv}        
	
	
	
	
	
	
	%
	%

\end{document}